\documentclass[]{bytedance_seed}

\usepackage[toc,page,header]{appendix}

\usepackage{minitoc}
\usepackage{lineno}
\usepackage{enumitem}  

\usepackage{latexsym}

\usepackage[T1]{fontenc}

\usepackage[utf8]{inputenc}

\usepackage{microtype}

\usepackage{inconsolata}

\usepackage{graphicx}

\usepackage{graphicx}
\usepackage{subcaption}
\usepackage{booktabs}
\usepackage{amsmath}
\usepackage{amssymb}
\usepackage{mathtools}
\usepackage{todonotes}
\usepackage{amsthm}
\usepackage{MnSymbol}
\usepackage{makecell}

\usepackage{bbding}
\usepackage{pifont}
\usepackage{color, xcolor}
\definecolor{skyblue}{rgb}{0.6, 0.98, 0.85}
\usepackage{soul}

\definecolor{codegreen}{rgb}{0,0.6,0}
\definecolor{codegray}{rgb}{0.5,0.5,0.5}
\definecolor{codepurple}{rgb}{0.58,0,0.82}
\definecolor{backcolour}{rgb}{0.95,0.95,0.92}

\usepackage{listings}

\lstdefinestyle{mystyle}{
    backgroundcolor=\color{backcolour},   
    commentstyle=\color{codegreen},
    keywordstyle=\color{blue},
    numberstyle=\tiny\color{codegray},
    stringstyle=\color{codepurple},
    basicstyle=\ttfamily\footnotesize,
    breakatwhitespace=false,         
    breaklines=true,                 
    captionpos=b,                    
    keepspaces=true,                 
    numbers=left,                    
    numbersep=5pt,                  
    showspaces=false,                
    showstringspaces=false,
    showtabs=false,                  
    tabsize=4
}

\newcommand{\pl}{\textsuperscript{+}}

\title{CodeContests\pl: High-Quality Test Case Generation for\\ Competitive Programming}

\author[1,2]{Zihan Wang}
\author[1]{Siyao Liu}
\author[1]{Yang Sun}
\author[2]{Hongyan Li}
\author[1]{Kai Shen}

\affiliation[1]{ByteDance Seed}
\affiliation[2]{Peking University}

\abstract{
Competitive programming, due to its high reasoning difficulty and precise correctness feedback, has become a key task for both training and evaluating the reasoning capabilities of large language models (LLMs). However, while a large amount of public problem data, such as problem statements and solutions, is available, the test cases of these problems are often difficult to obtain. Therefore, test case generation is a necessary task for building large-scale datasets, and the quality of the test cases directly determines the accuracy of the evaluation. In this paper, we introduce an LLM-based agent system that creates high-quality test cases for competitive programming problems. We apply this system to the CodeContests dataset and propose a new version with improved test cases, named CodeContests\pl. We evaluated the quality of test cases in CodeContestsPlus. First, we used 1.72 million submissions with pass/fail labels to examine the accuracy of these test cases in evaluation. The results indicated that CodeContests\pl\; achieves significantly higher accuracy than CodeContests, particularly with a notably higher True Positive Rate (TPR). Subsequently, our experiments in LLM Reinforcement Learning (RL) further confirmed that improvements in test case quality yield considerable advantages for RL.
}

\date{\today}
\correspondence{Kai Shen at \email{shen.kai@bytedance.com}}

\checkdata[Project Page]{\url{https://huggingface.co/datasets/ByteDance-Seed/Code-Contests-Plus}}

\begin{document}
\maketitle

\section{Introduction}
\label{sec:intro}

Competitive programming is widely recognized as an important benchmark for evaluating the reasoning and coding capabilities of LLMs \cite{o1}. Solving complex competitive programming problems requires strong reasoning capabilities, as well as mastery of a wide range of algorithms, data structures, and mathematical knowledge. More importantly, competitive programming problems are objectively verifiable tasks; thus, they are not only widely used for benchmarks, but they can also provide accurate rewards for reinforcement learning and serve as a vital data foundation for training large reasoning models \cite{seedthinking, deepseekr1}.

Existing open-source competitive programming datasets usually collect problems from competition platforms like CodeForces \cite{cf}, LeetCode, and AtCoder. However, these competition platforms do not publicly release their test cases. Consequently, although large amounts of problems with statements and solutions are publicly available, the lack of test cases prevents these problems from being effectively constructed into RL training datasets.

Consequently, existing datasets primarily utilize their own created test cases, rather than fully using the official ones. Commonly used automated test case generation methods include mutation, as well as some LLM-based methods. However, the quality of these automatically generated test cases still falls far short of those designed by professional human problem setters. Specifically, the main gaps lie in the following two aspects:

\begin{figure}
    \centering
    \includegraphics[width=0.5\linewidth]{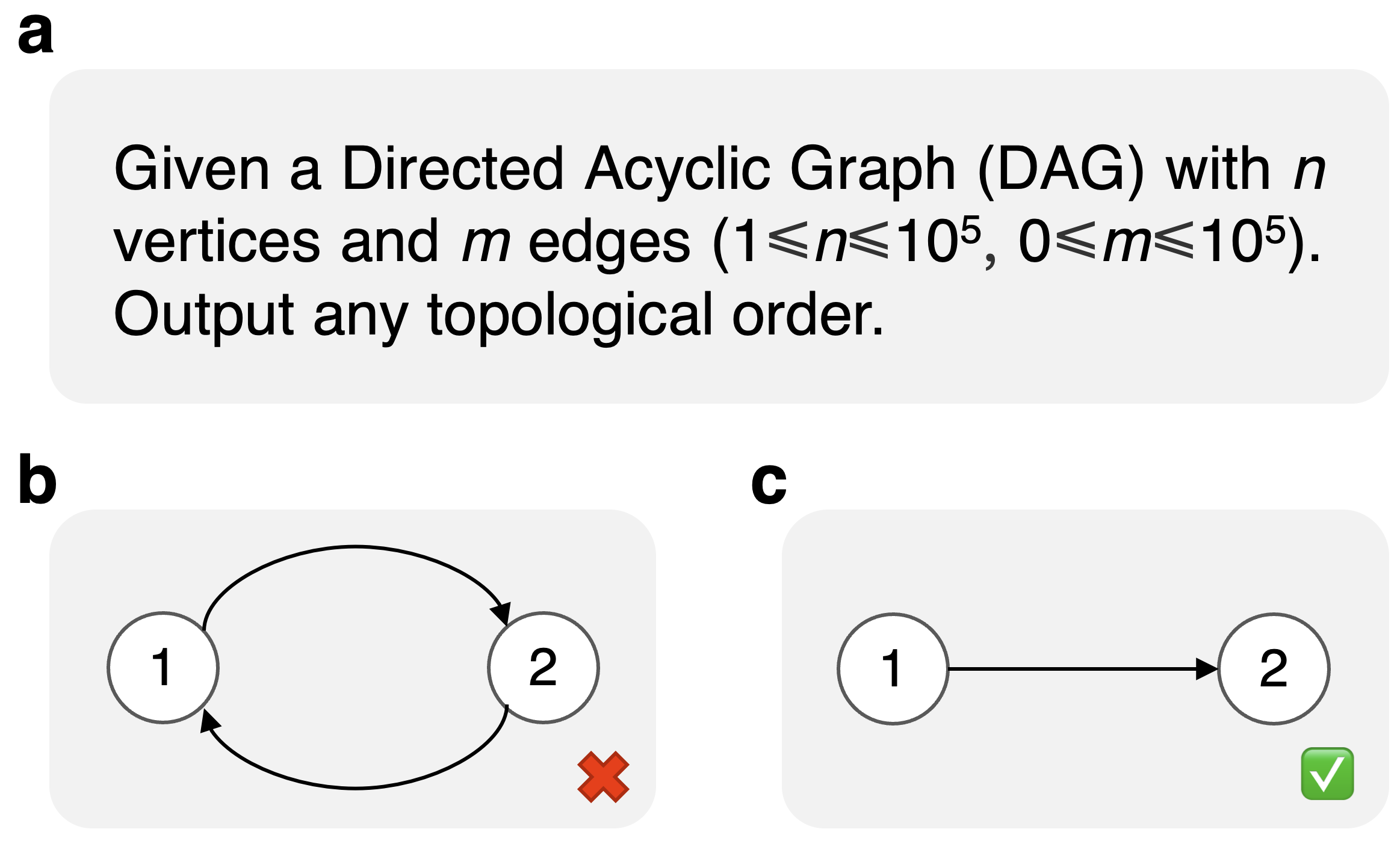}
    \caption{Competitive programming problems typically impose constraints on the input. (a) A simple example of a topological sort problem, which requires the input to be a Directed Acyclic Graph (DAG) and specifies limits on its size. (b) An invalid input, as the graph contains a cycle, which means no topological sort exists. (c) A valid input.}
    \label{fig:intro}
\end{figure}

\begin{itemize}
    \item \textbf{Limited coverage.} Some methods, such as Mutation, can only blindly construct large amounts of random data. They struggle to generate tricky cases and corner cases that require deep algorithmic understanding to discover, making it hard to cover deep-level, complex boundary conditions or special situations within the problem's logic. Additionally, some methods, like directly outputting test cases via LLMs, are often unable to generate large-scale test cases. Therefore, they cannot identify incorrect solutions that are logically correct but fail due to excessively high time or memory complexity. Limited coverage may lead to false positives, meaning that incorrect solutions might be judged as correct.
    \item \textbf{Incorrect test cases.} As shown in Fig. \ref{fig:intro}, a typical programming problem usually imposes constraints on the test cases themselves. Existing methods struggle to ensure that the generated test cases can satisfy these constraints, and incorrect test cases can simultaneously lead to both false positives and false negatives. To the best of our knowledge, the issue of incorrect test cases has not received attention in previous research. We examined the test cases in the CodeContests dataset and found that incorrect test cases are one of the main causes of inaccurate evaluation.
\end{itemize}

In this paper, we propose an LLM-based agent system for constructing test cases for programming problems that have more comprehensive coverage and better correctness. This will allow for further improvement in the quality and scale of code RL datasets. Specifically, to address the above two limitations, we propose the following solutions:

\begin{itemize}
    \item \textbf{An Agent for Test Case Generation.} We designed a Generator agent that writes a generator program for each problem to specifically construct diverse test cases, including random data, corner cases, and tricky cases, thereby fully testing various possible solutions and potential error patterns, as well as examining the efficiency of the algorithm through large test cases. This generator program can be run any number of times with different random seeds, thus obtaining any number of test cases, further improving coverage.
    \item \textbf{An Agent for Test Case Validation.} Although this generator agent consciously attempts to satisfy the constraints in the problem, it still has a noticeable probability of making mistakes. Therefore, we designed a validator agent. This agent writes a validator program to check whether the input of the generated test cases satisfies all the constraints in the problem. Incorrect test cases and the specific reasons for the errors will be fed back to the generator for revision until all test cases satisfy the conditions.
\end{itemize}

The contributions of this paper is summarized as follows:

    \begin{enumerate}
        \item \textbf{An LLM-Based Agent System for Test Case Construction.} We propose the Generator-Validator (G-V) agent system, the first LLM agent system designed for constructing high-quality test cases for competitive programming problems.
        \item \textbf{A Code Dataset with Verified Test Cases.} Using the G-V agent system, we create CodeContests\pl, the first competition-level code dataset with \textbf{verified} test cases. We verify the coverage and correctness of the test cases by evaluating the true positive rate (TPR) and true negative rate (TNR) of each problem using 1.72 million labelled solutions. CodeContests\pl share the same problem set with CodeContests, but replacing test cases with those generated by our G-V agent system.
        \item \textbf{Study.} Comparing under the same TPR and TNR thresholds, CodeContests\pl can yield twice the number of effective problems compared to CodeContests, thus validating that the test case quality of CodeContests\pl is significantly better than that of CodeContests. We trained a 32B reasoning model using RL separately with CodeContests\pl and CodeContests and observed a clear advantage for CodeContests\pl during the training process.
    \end{enumerate}

\begin{table*}
\caption{Comparison between CodeContests\pl\;and other code datasets and benchmarks.}
\footnotesize
    \centering
    \begin{tabular}{lccccl}
    \toprule
         & Type & \# Problems & \thead{Problem\\Difficulty}  & \thead{Customized\\Checker}  & \thead{How Test Cases\\Are Constructed?}\\
    \midrule
    
    MBPP \cite{mbpp} & Benchmark & 974 & $\bigstar$ &  \ding{55} & Handcrafted \\
    HumanEval \cite{humaneval} & Benchmark & 164 & $\bigstar$  &\ding{55} & Handcrafted\\
    USACO \cite{usaco}  & Benchmark & 307 & $\bigstar\bigstar$  &\ding{55} & Publicly accessible\\
    LiveCodeBench \cite{livecodebench} & Benchmark & 1055 & $\bigstar\bigstar$   &\ding{55} & Semi-automatic \\
    APPS \cite{apps} & Train & 10000 & $\bigstar\bigstar\bigstar$  &\ding{55} & Crawled \\
    CodeContests \cite{alphacode} & Train & 13610 & $\bigstar\bigstar\bigstar$   &\ding{55} & Mutation \\
    TACO \cite{taco}  & Train & 26433 & $\bigstar\bigstar\bigstar$   &\ding{55} & Output by LLM \\
    \midrule
    CodeContests\pl (Ours)  & Train & 11690 & $\bigstar\bigstar\bigstar$  &\ding{51} & G-V agent system \\
    \bottomrule
    \end{tabular}
    
    \label{tab:related_work}
\end{table*}
\section{Related Work}
\label{sec:related}

Since most competitive programming platforms do not disclose their test cases, constructing test cases is one of the primary bottlenecks in building code datasets. The test case generation methods currently employed by existing code datasets can generally be categorized into three types: manual generation, mutation-based generation, and LLM-based generation.

\textbf{Manual.} Representative works that use manually constructed test cases include MBPP \cite{mbpp}, HumanEval \cite{humaneval}, and LiveCodeBench \cite{livecodebench}. There are slight differences among these three: the test cases in MBPP and HumanEval are handcrafted, resulting in a smaller quantity and insufficient coverage. In contrast, for a portion of the problems in LiveCodeBench, test cases are constructed by human experts specifically targeting the problem characteristics, leading to better coverage.

The common drawbacks of manually constructing test cases are their high cost, lack of automation, and difficulty in scaling up. Therefore, such methods are only suitable for building small-scale evaluation sets and are too costly to use for constructing large-scale training sets.

\textbf{Mutation-Based.} \citet{evalplus} identified the issue of high False Positive Rates (FPR) in MBPP and HumanEval due to their small number of test cases. They proposed ``Type-aware input mutation" to generate new test cases by recombining a few existing test cases. A similar mutation approach was also employed to construct the CodeContests \cite{alphacode} dataset. The advantage of mutation methods lies in their complete automation, allowing for the generation of a large volume of test cases. However, their limitation is that if a problem involves complex constraints, mutation often fails to satisfy these constraints, thereby introducing incorrect test cases and leading to a high False Negative Rate (FNR).

\textbf{LLM-based.} Since competitive programming problems often involve complex constraints on test input, some approaches turn to LLMs to handle this. TACO \cite{taco}, for instance, uses LLMs to directly output test input. However, while LLMs can access the problem description and understand the constraints to some extent, they are not guaranteed to output an input that satisfies these constraints. Furthermore, this method can only construct small test cases and is limited by the context window size, preventing it from outputting large test cases. For example, an LLM cannot directly output a graph containing one million vertices.

\section{The Generator-Validator Agent System}
\label{sec:method}

\begin{figure*}[htbp]
    \centering
    \includegraphics[trim=0 0 0 8, clip,width=1\linewidth]{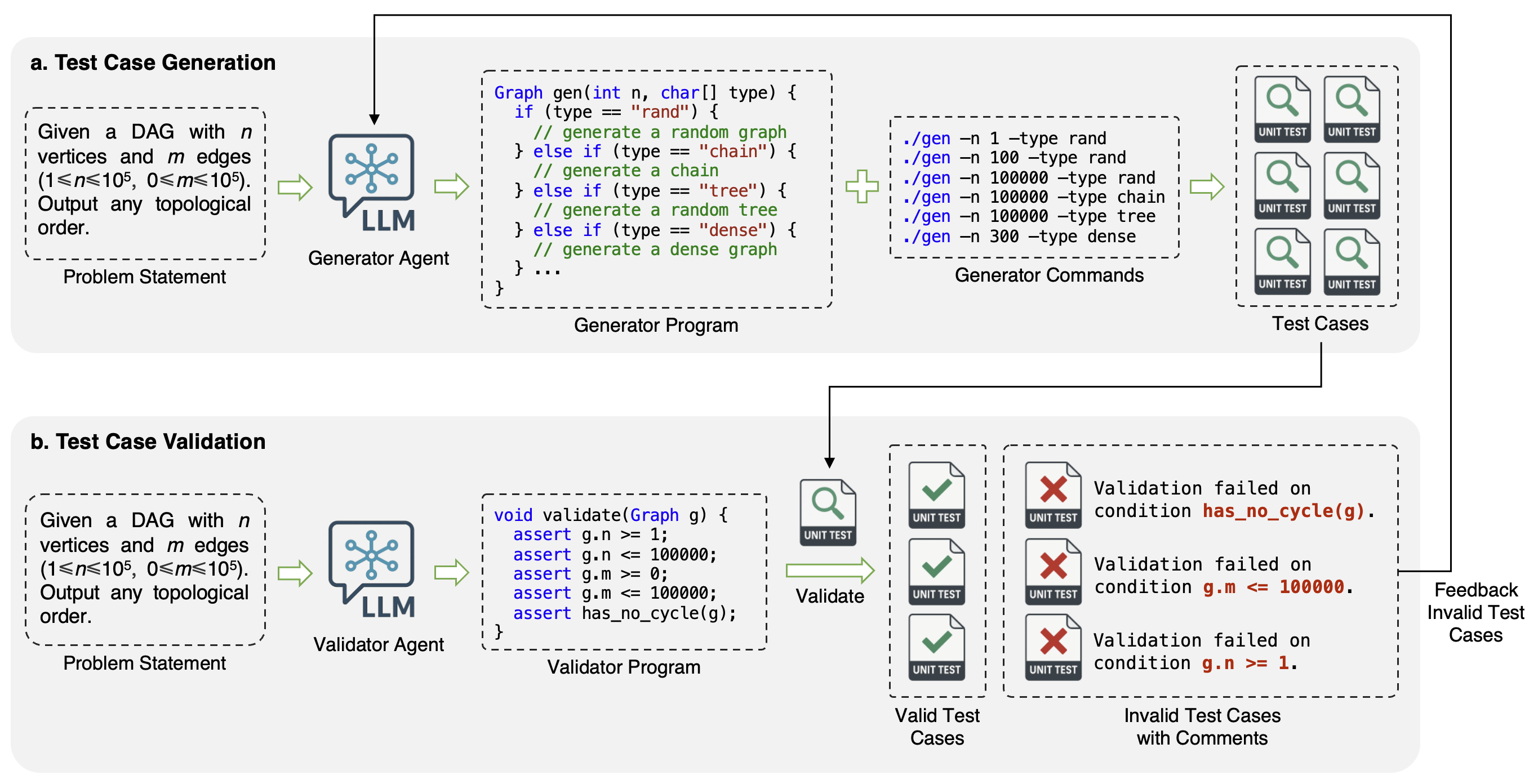}
    \caption{\textbf{Generator-Validator Agents Pipeline.} (a) The Generator Agent writes a generator program and generator commands to produce test cases. (b) The Validator Agent writes a validator program to check if the test cases satisfy all the constraints stated in the problem. Test cases that fail validation, along with specific comments provided by the validator program, are then fed back to the Generator Agent for revision. }
    \label{fig:workflow}
\end{figure*}

To simultaneously address the challenges of correctness and coverage in test case construction, we propose a Generator-Validator agent system. The Generator agent is an LLM-based agent that writes test input generators based on the problem description. The Validator agent is also an LLM agent that writes test input validators to supervise the Generator agent, ensuring that the test input generated by the Generator agent conforms to the problem constraints. This agent system can automatically build a large-scale and rigorous dataset for LLM training by leveraging publicly available data such as problem statements and ground truth solutions. An overview of the workflow is presented in Fig. \ref{fig:workflow}. Specific implementation details for the generator and validator are presented below.

\subsection{Generator}
\label{sec:generator}

\textbf{Generator Program.} The generator program accepts input data conditions, such as data size and characteristics, via command-line arguments. It then generates a piece of input data that conforms to these specified conditions. This generated input data is then fed into the ground truth solution to produce outputs and hence form a complete set of test cases. The advantage of this design is that a single generator program, combined with varying command-line arguments, can generate a diverse range of test cases. A generator demo is presented in Appendix \ref{sec:demogenerator}.

\textbf{Agent Workflow.} Initially, the Generator Agent is given a problem statement. It is then instructed to read the statement carefully, and identify, and summarize the constraints of the input data. Subsequently, the Generator Agent will analyze the problem to anticipate potential mistakes contestants might make and identify possible corner cases. Based on this analysis, it will design targeted adversarial test cases. Finally, the Generator Agent will synthesize all of this information to produce a compliant generator program. Since it is up to the agent to determine which command-line arguments the generator program needs to receive, we also require the agent to provide approximately 20 commands. These commands should cover a range of data sizes, from small to large, and all relevant special types. We will then execute these commands to finally obtain the test input.

\textbf{Details.} The generator agent is instructed to use \texttt{testlib} \cite{testlib}, a tool library developed by Codeforces for contest problem setters, to implement the generator. \texttt{testlib} provides some useful utility functions, such as random number generators and command-line argument parsing tools. Building upon testlib, we have developed a more LLM-friendly version, reducing the difficulty of LLM usage and the likelihood of compilation errors and hallucinations. The use of \texttt{testlib} helps regularize the behavior of LLM-written generators. For instance, we enforce the use of \texttt{testlib}'s random number generator, rather than using the C++ standard library's random facilities. This is to ensure random number consistency, i.e., the same command and the same generator, even on different platforms, will produce identical test cases.

\textbf{Scalability.} Some commands provided by the agent can generate not only a single test case but also an arbitrary number of test cases by altering the random seed. The random seed for the generator program is calculated based on the hash of the command. The agent is not allowed to modify the random seed in the generator program, ensuring that the same command always produces identical test cases. We can set different random seeds by appending an irrelevant label to the end of the command. This label will not be parsed, and therefore will not affect the behavior of the generator program. In this way, we can flexibly change random seeds and ensure consistency across different platforms at the same time. The number of test cases can be adjusted according to actual needs. For example, the number could be reduced during training to minimize evaluation time overhead and improve training efficiency. When used for benchmarking, the number of test cases may be increased.

\begin{figure}
    \centering
    \includegraphics[width=0.5\linewidth]{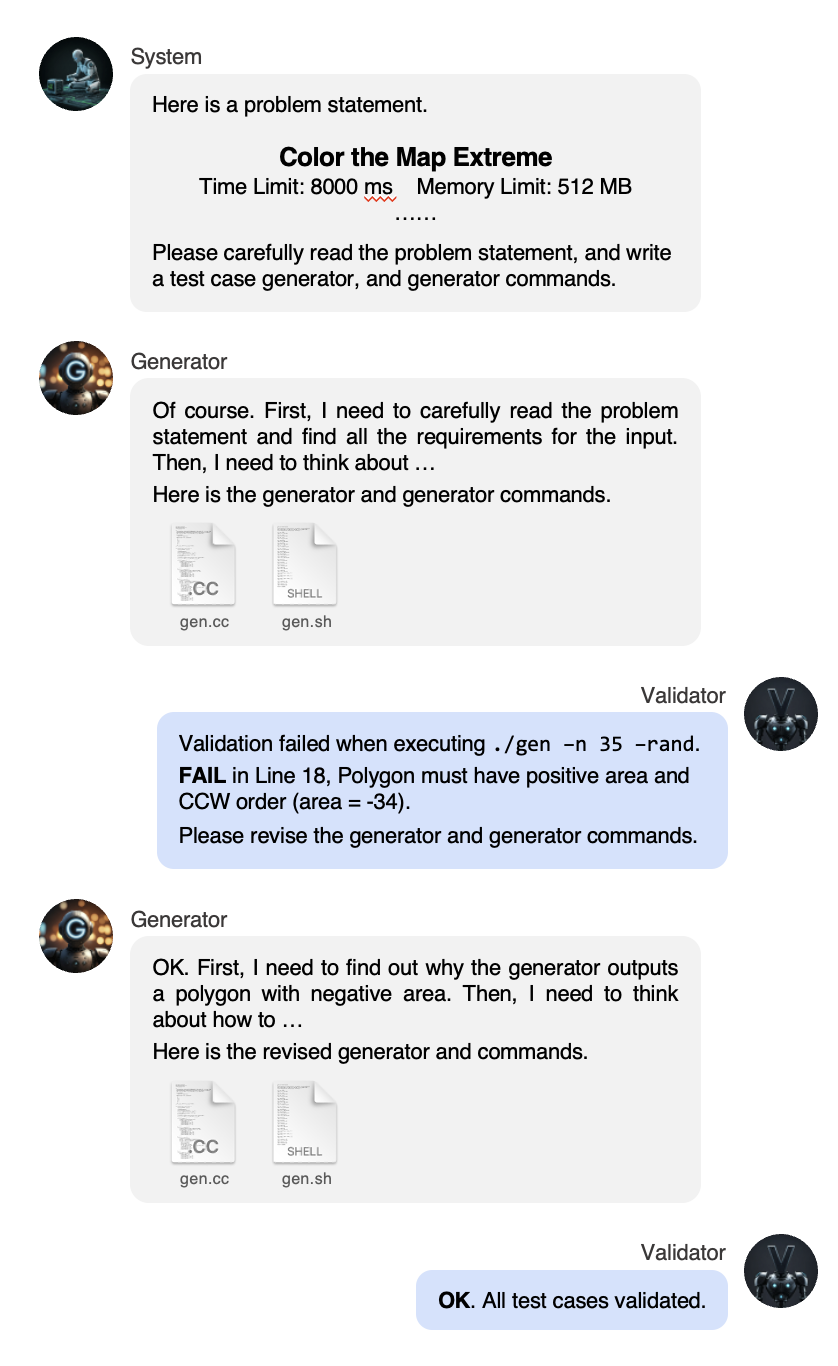}
    \caption{An illustration of the reflection process of Geneartor Agent.}
    \label{fig:reflection}
\end{figure}

\textbf{Supervision.} The generator agent is very likely to make mistakes while writing the generator program. Although the agent may recognize some constraints in the problem description, it can still miss specific details either partially or entirely due to limited attention or imperfect comprehension. Therefore, a supervision mechanism is necessary to help the generator agent identify and correct errors. We use a validator to check if the test input satisfies all the constraints specified in the problem statement. If errors are found, it provides specific error locations and causes. These error reports are then fed back to the generator agent. Subsequently, the agent reflects and corrects the issue, providing a revised generator program and commands.  Additionally, specific error messages for other potential errors, like compilation errors or generator timeouts, are also provided to the agent for further revision. An example of the supervision and reflection procedure is shown in Fig. \ref{fig:reflection}. The implementation details of the validator will be discussed in Section \ref{sec:validator}.

\subsection{Validator}
\label{sec:validator}

\textbf{Validator Program.} Generating test cases is a very intricate task, so even professional competitive programming problem setters can make mistakes sometimes due to oversight. For example, in the ACM ICPC World Finals 2007, Problem J was found to have an incorrect test case due to an error by the problem setter. As mentioned above, the validator plays an even more crucial role for the generator agent; it not only provides a double check on the correctness of test cases but also provides important supervision information to help the generator agent reflect and correct its errors. A validator is a program that takes one input data as its input and determines whether this input data satisfies all the constraints of the problem. If errors exist, the validator outputs exactly which constraints were violated. For errors where the location can be specifically identified, the validator also provides the error location, for example, the line number in the input data where the error was found. 
An example of a validator is presented in Appendix \ref{sec:demovalidator}.

\textbf{Agent Workflow.} Initially, the validator agent is provided with a problem statement. Next, the agent is required to carefully read the problem statement, identify all input data constraints, including data ranges, format requirements, and structural constraints, and summarize them. Finally, the agent will write a validator program to check these constraints.

\textbf{Supervision.} While the probability of a validator agent making mistakes is much lower than that of a generator agent, errors are still possible. Based on our observations, common errors fall into two categories: The first is where the agent correctly understands the constraints but makes mistakes while writing the validator program. The second is where the agent overlooks some of the constraints stated in the problem. Errors of the first type can lead to the program failing to compile or run correctly, or causing valid input data to fail the validation. Therefore, for the first type of error, we feed the sample inputs from the problem statement to the validator and check whether these data pass the validator's checks. If not, it indicates an error in the validator. In this case, both the sample data and the validator's output are fed back to the validator agent, which then reflects and makes revisions based on the feedback. Additionally, the validator receives specific error messages for common issues like compilation failures, runtime errors, and timeouts. This supervision mechanism could detect most of the errors of the agent. Unfortunately, we still lack an automatic supervision method to address the second type of error, which can still result in a small number of incorrect data being generated. Detailed statistics and case studies will be presented in Section \ref{sec:evaluation} below.

\section{CodeContests\pl: A Competitive Coding Dataset with Verified Test Cases}
\label{sec:dataset}

\begin{table}
    \centering
    \caption{Comparison between CodeContests and CodeContests\pl}
    \small
    \begin{tabular}{ccc}
    \toprule
         & CodeContests & CodeContests\pl\\
         \midrule
        Problem Count & 13610 & 11690\\
        Average Tests & 101 & 25/44/62/80/98/$\infty$\\
        Validation Pass Rate & 67.1\% & 100\%\\
        Generator & \ding{55} & \ding{51} \\
        Validator & \ding{55} & \ding{51} \\
        Checker & \ding{55} &  \ding{51}\\
    \bottomrule
    \end{tabular}
    
    \label{tab:ccccplus}
\end{table}

CodeContests \cite{alphacode} stands as one of the largest and most widely recognized competitive coding datasets. It collects a large number of problems, authentic contestant submission records, and generates numerous additional test cases through mutation.  However, test cases generated through mutation are often of low quality and may yield unreliable evaluation results, such as misclassifying incorrect solutions as correct and vice versa. In this Section, we present our methodology for constructing an enhanced dataset, CodeContests\pl, by building upon the original CodeContests.
Section \ref{sec:cleaning} outlines the data cleaning procedures we implemented.
Section \ref{sec:gentests} details how we utilized the G-V Agent system to generate higher-quality test cases for CodeContests\pl.
Section \ref{sec:intro_checker} describes the development of customized checkers for problems that accept multiple valid solutions.

Following this, we compare the quality of CodeContests\pl and CodeContests across two primary dimensions.
First, in Section \ref{sec:evaluation}, we verify the quality of the test cases of both datasets. More specifically, we utilized 1.72 million authentic contestant submissions, which comprise both correct and incorrect ones, to assess the performance of both datasets in discriminating between correct and incorrect solutions.
Then, in Section \ref{sec:rl}, we employed each dataset to train a 32B LLM through DAPO \cite{yu2025dapo}, to evaluate the impact of dataset quality on RL training efficacy.

\subsection{Data Cleaning}
\label{sec:cleaning}
We examined the problems in the CodeContests dataset and identified some that were either incorrect or unsuitable for training. We then cleaned up these problems. Specifically, we removed the following types of problems: (1) problems without problem statements, (2) interactive problems, (3) problems without correct submissions, (4) problems involving file input/output (5) special problems, such as April Fools' Day problems, (6) problems that require images for proper understanding, (7) problems with crawling errors, (8) other low-quality problems, e.g., problems which lack data ranges, have unclear requirements, or contain incorrect sample formats. After cleaning, the total problem count was reduced from 13,610 to 11,690.

It should be noted that due to our additional filtering work, CodeContests and CodeContests\pl differ in the number of problems. In the experiments in Sections \ref{sec:evaluation} and \ref{sec:rl}, our focus is solely on the impact of test case quality. Therefore, we selected the same subset of problems for CodeContests as well, ensuring that the set of problems used for both datasets in these experiments is identical.

\subsection{Generated Test Cases using G-V Agent System}
\label{sec:gentests}

In CodeContests\pl, we have replaced all original test cases from CodeContests with generators and validators produced by the G-V Agent system. Furthermore, we offer two methods for utilizing these test cases:

\textbf{Dynamic Generation.} Using the provided generators and validators to produce test cases. This approach offers greater flexibility, allowing for the generation of an arbitrary number of test cases based on specific needs and computational budgets. As detailed in Section \ref{sec:generator}, we have ensured random consistency, guaranteeing that using the same generator and commands will produce identical test cases across different platforms. To facilitate the execution of these generators and validators and enable automated evaluation, we open-source SandboxFusion\footnote{\url{https://github.com/bytedance/SandboxFusion}}, which supports over 20 programming languages, including C++, Java, Python, Rust, Go, and more than 10 open-source datasets, such as MBPP, HumanEval, CodeContests.

\textbf{Pre-Processed Test Cases.} We release five versions of pre-generated test cases, labeled CodeContests\pl1x through CodeContests\pl5x. These correspond to running each generator command 1 to 5 times, respectively, each time with a different random seed. The 1x version contains an average of 25 test cases per problem, while the 2x, 3x, 4x, and 5x versions contain averages of 44, 62, 80, and 98 test cases per problem, respectively. Using these pre-generated test sets offers easier compatibility with other sandboxes and evaluation environments. The statistics of CodeContests\pl and CodeContests are presented in Table \ref{tab:ccccplus}. All the pre-processed test cases have passed the validators. We used validators to check the correctness of all 1.18 million generated test cases in CodeContests. Only 0.79 million of these passed validation, which is 67.1\%.

\subsection{Customized Checkers for Multiple-Answer Problems}
\label{sec:intro_checker}

Problems with multiple valid solutions are common in programming contests. For such problems, multiple distinct outputs can be considered correct for the same input. For instance, as illustrated by the examples in Fig. \ref{fig:intro} and Fig. \ref{fig:workflow}, a DAG can have multiple valid topological sorts, and any one of these is an acceptable output. Furthermore, the number of correct solutions can be vast, potentially even infinite. For example, a DAG with $n$ nodes can have up to $n!$ distinct topological sorts, making it infeasible to enumerate and store all of them.

Previous datasets, such as CodeContests, have collected numerous problems with multiple solutions but lack customized judging logic for them. We have developed a Checker Agent that provides customized checker programs for all problems. A checker program is designed to determine if a code's output is correct for a given input. Taking the topological sort problem as an example, its checker program would read the input graph and the submitted code's output, then verify if this output constitutes a valid topological order.

Due to space constraints, the implementation details of the Checker Agent are provided in the Appendix \ref{sec:checker}.

\subsection{Test Case Quality Verification}
\label{sec:evaluation}

\begin{figure}
	\centering
\begin{subfigure}[b]{5.2cm}
		\includegraphics[trim=75 50 90 90,clip,width=\linewidth]{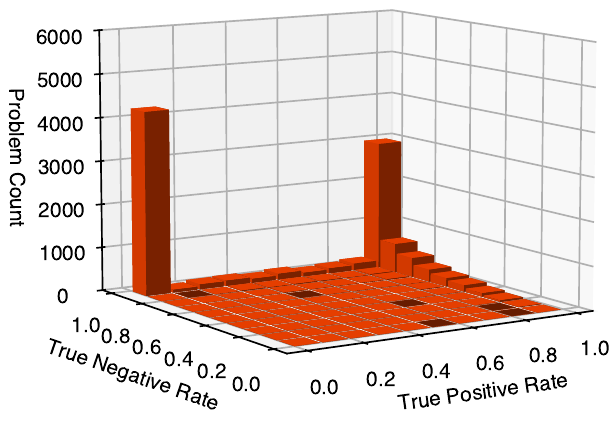}
		\caption{CodeContests}
		\label{fig:cc}
	\end{subfigure}
	\begin{subfigure}[b]{5.2cm}
		\includegraphics[trim=75 50 90 90,clip,width=\linewidth]{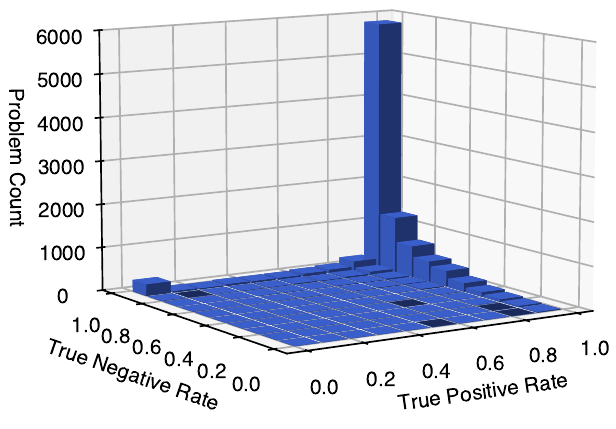}
		\caption{CodeContests\pl\;5x}
		\label{fig:5x}
	\end{subfigure}
    \begin{subfigure}[b]{5.2cm}
		\includegraphics[trim=5 10 20 40,clip,width=1\linewidth]{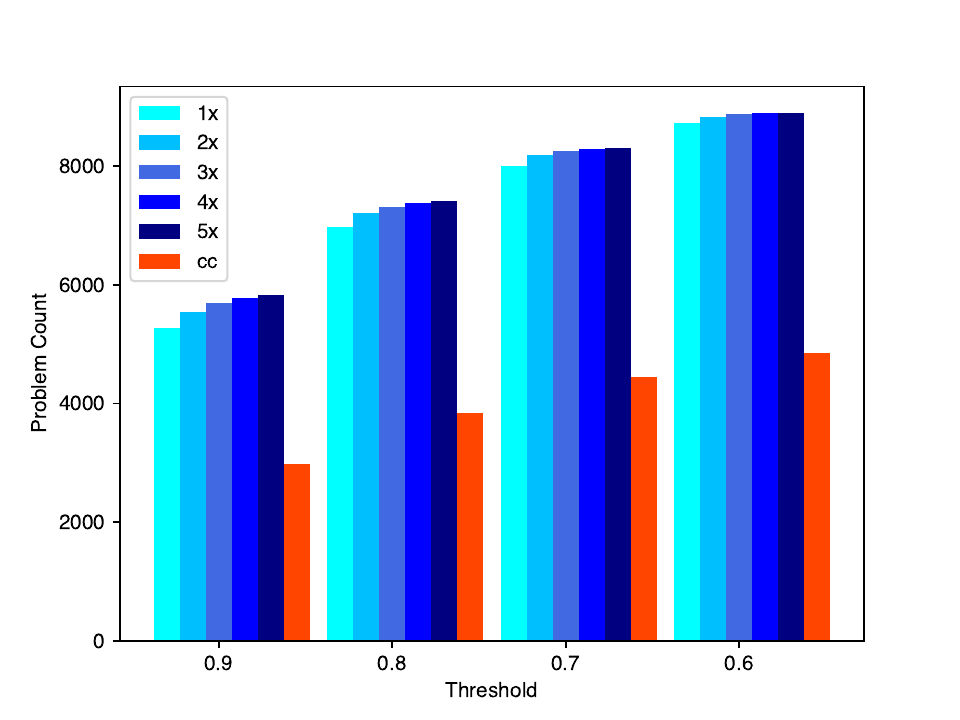}
		\caption{\#Problems v.s. Threshold}
		\label{fig:threshold}
	\end{subfigure}
\caption{The histogram of the TPR and TNR of selected problems from (a) CodeContests and (b) CodeContests\pl, and (c) the number of qualified problems with TPR and TNR greater than a threshold in CodeContests\pl (blue ones, ours) and CodeContests (red).}
\label{fig:tptn}
\end{figure}

\textbf{Task.} Test cases serve to determine the correctness of a given code. Therefore, we treat code evaluation as a binary classification problem. This approach allows us to evaluate the test cases themselves as binary classifiers, assessing their ability to distinguish between correct and incorrect solutions accurately. By doing so, we can objectively and rigorously evaluate the quality of test cases within a code dataset, including their coverage and correctness. To our knowledge, CodeContests\pl\; is the first dataset project to conduct such rigorous validation of test case accuracy. We believe this level of evaluation is crucial for establishing a dataset's trustworthiness.

\textbf{Data.} CodeContests has collected tens of millions of submissions, each with a ``correct''/``incorrect'' label. We sampled 100 positive samples (correct submissions) and 100 negative samples (incorrect submissions) for each problem. For problems without sufficient samples, we included as many available submissions as possible. In this way, we selected 10,166 problems that contained at least 10 positive and negative samples, as well as a corresponding 1.72 million submissions. We used these problems and samples to evaluate the accuracy of the test cases in CodeContests\pl\; and CodeContests.

\textbf{Engineering.} Since each submission needs to be evaluated on approximately 200 test cases on average, the total number of program executions is more than 300 million. We implemented a cloud architecture for such large-scale evaluation, running on a cluster with 25,000 CPU cores and 70 TB memory, and completed the experiment on this platform. Some engineering details are presented in Appendix \ref{sec:cloud}.

\textbf{Metric.} We use True Positive Rate (TPR) and True Negative Rate (TNR) to quantitatively assess accuracy. TPR measures the ability of test cases to correctly classify positive instances (correct solutions), thus reflecting the test case's correctness. This is because if a test case satisfies all the problem's constraints, a correct solution should not be misclassified as incorrect. Therefore, a low TPR primarily indicates that the test case itself is flawed. TNR measures the ability of test cases to correctly classify negative instances (incorrect solutions) as incorrect, thereby primarily reflecting the test case's coverage. Overly simplistic test cases may fail to identify errors in incorrect code, leading to false positives (i.e., incorrect solutions being deemed correct).

\textbf{Results.} We calculated the TPR and TNR for each problem in CodeContests and CodeContests\pl 5x and plotted histograms, which are shown in Fig \ref{fig:tptn}. From the results, it is clearly observed that in CodeContests, there are over 4000 problems with $\text{TPR}\le 0.1$ and $\text{TNR}\ge 0.9$, indicating that these problems incorrectly classify almost all correct submissions as incorrect, making these problems practically impossible to solve and cannot be used in training. We analyzed these problems and identified two primary reasons for this phenomenon. First, CodeContests includes a large number of incorrect test cases, causing program outputs to be meaningless or causing programs to fail to run properly. Second, CodeContests does not provide custom checkers for multi-solution problems. In contrast, our proposed Agent System can better ensure the correctness of test cases and provide custom checkers for multi-solution problems, so similar phenomena are not as prominent.

\begin{figure*}

	\centering

\begin{subfigure}[b]{5cm}
    \includegraphics[trim=75 50 90 90,clip,width=5cm]{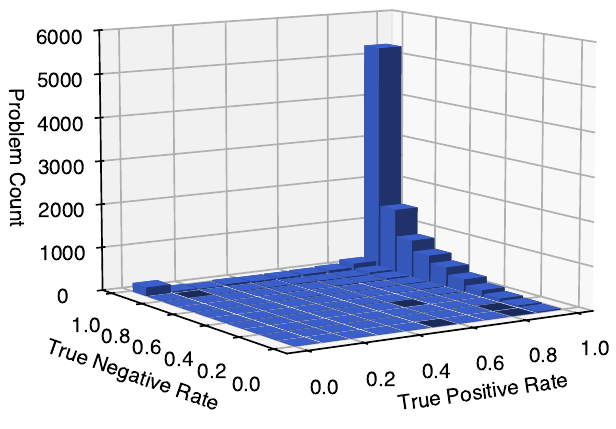}
    \caption{CodeContests\pl\;1x}
	\label{fig:1x}
\end{subfigure}
\begin{subfigure}[b]{5cm}
    \includegraphics[trim=75 50 90 90,clip,width=5cm]{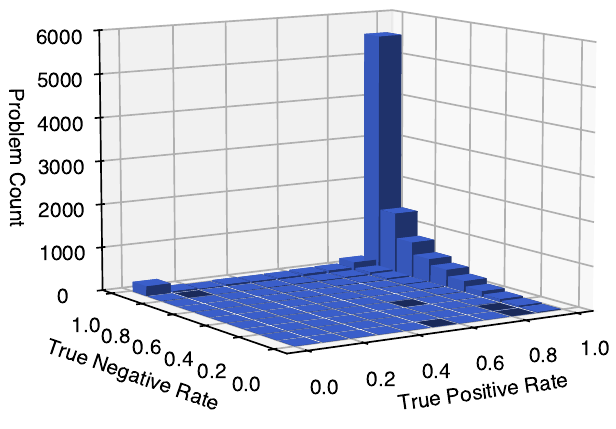}
    \caption{CodeContests\pl\;2x}
	\label{fig:2x}
\end{subfigure}
\begin{subfigure}[b]{5cm}
    \includegraphics[trim=75 50 90 90,clip,width=5cm]{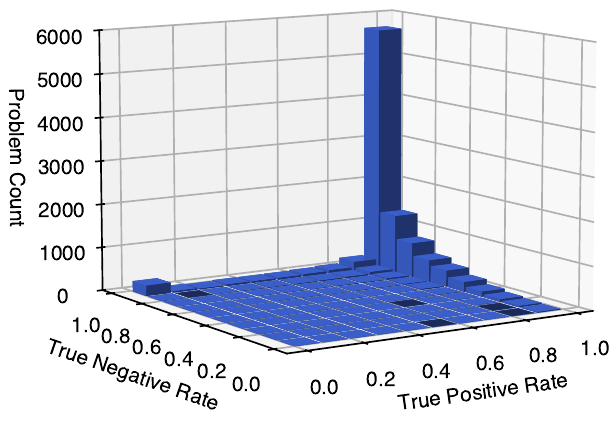}
    \caption{CodeContests\pl\;3x}
	\label{fig:3x}
\end{subfigure}
\begin{subfigure}[b]{5cm}
    \includegraphics[trim=75 50 90 90,clip,width=5cm]{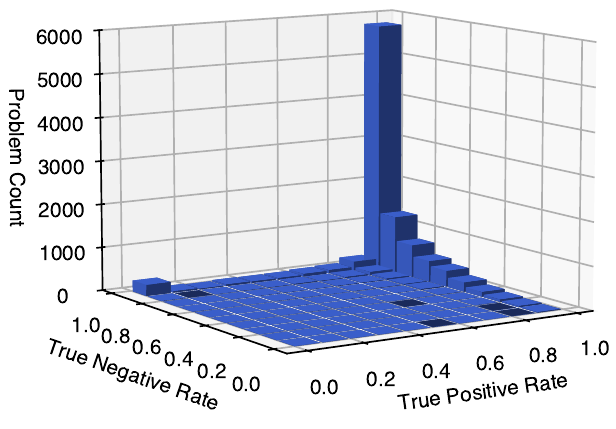}
    \caption{CodeContests\pl\;4x}
	\label{fig:4x}
\end{subfigure}
\begin{subfigure}[b]{5cm}
    \includegraphics[trim=75 50 90 90,clip,width=5cm]{figs/tp_tn_5x.pdf}
    \caption{CodeContests\pl\;5x}
	\label{fig:5x}
\end{subfigure}
\begin{subfigure}[b]{5cm}
    \includegraphics[trim=75 50 90 90,clip,width=5cm]{figs/tp_tn_cc.pdf}
    \caption{CodeContests}
	\label{fig:cc}
\end{subfigure}

\caption{The histogram of the TPR, TNR of problems in (a)-(e) CodeContests\pl\;1x-5x (ours) and (f) CodeContests.}
\label{fig:tptn_complete}
\end{figure*}

Furthermore, we can observe that as the number of test cases increases (from 1x to 5x), the overall evaluation accuracy improves. This is reflected in Fig. \ref{fig:tptn_complete} by the gradual increase in the number of problems with $\text{TPR\&TNR} \geq 0.9$. To better demonstrate this change, we counted the number of problems with $\text{TPR\&TNR}$ greater than a given threshold, and the results are shown in Fig. \ref{fig:threshold}. The results show that, at various thresholds, the number of qualified problems in CodeContests\pl\;increases as the number of test cases rises. In particular, the number of qualified problems in CodeContests\pl\;5x is almost twice that of CodeContests. Even with only one-quarter of the test cases as compared to CodeContests, the 1x version of CodeContests\pl\;has a significantly higher evaluation accuracy and yields over 80\% more qualified problems than CodeContests.

This large-scale verification effort, as described above, allows us to further refine our selection by identifying problems with high-quality test cases and excluding those with problematic ones. Consequently, we have selected problems achieving both $\text{TPR\&TNR}\ge 0.9$ to form a distilled subset, which we designate CodeContests\pl HQ. In the next Section, we will use both CodeContests\pl HQ and the original CodeContests for RL training to demonstrate that higher-quality test cases yield significant benefits for the training process. Within our dataset, we provide the TPR and TNR for each problem, enabling users to set appropriate thresholds for quality-based filtering according to their specific requirements.

\begin{table}
\caption{Pass@1 on LiveCodeBench across difficulty levels. CodeContests\pl achieves consistent gains over CodeContests.}
\centering
\small  
\setlength{\tabcolsep}{3.5pt}  
\begin{tabular}{lcccc}
\toprule
\textbf{Dataset} & \textbf{Easy} & \textbf{Medium} & \textbf{Hard} & \textbf{All} \\
\midrule
CodeContests    & 0.958 & 0.786 & 0.329 & 0.622 \\
CodeContests\pl HQ  & \textbf{0.965} & \textbf{0.812} & \textbf{0.340} & \textbf{0.637} \\
\bottomrule
\end{tabular}
\vspace{2pt}
\label{tab:livecodebench-compact}
\end{table}

\subsection{Test Case Quality Matters in RL Training}
\label{sec:rl}

To investigate the effect of unit test coverage on code generation performance, we conduct a controlled ablation study using reinforcement learning (RL) based on the PPO-style training paradigm~\cite{shojaee2023ppocoder}.

\textbf{Benchmark.} Our experiments are performed on the LiveCodeBench benchmark~\cite{livecodebench}, which evaluates code generation models across varying difficulty levels: \textit{Easy}, \textit{Medium}, and \textit{Hard}. The time window is Aug 2024 - Feb 2025.
We use avg@15 as the performance metric, which is the average of pass@1 from 15 independent responses.

\textbf{Cold Start.} We initialize our policy from Qwen2.5-32B~\cite{qwen2024technical}, a reasoning-optimized large language model. To enhance its zero-shot reasoning capabilities, we further perform supervised fine-tuning using a curated cold-start reasoning dataset.

\textbf{Optimization Objective.}
We adopt the \textit{Decoupled Clip and Dynamic Sampling Policy Optimization (DAPO)} objective~\cite{yu2025dapo} to enhance policy learning via group-based advantage estimation.
Given an input query \( q \sim P(Q) \), a group of \( G \) candidate outputs \( \{o_{i, t}\}_{i=1}^G \) is sampled from the policy \( \pi_{\theta_{\text{old}}}(O \mid q) \) using a dynamic sampling strategy.
The DAPO objective is defined as:
\begin{equation}
\begin{aligned}
\mathcal{J}_{\text{DAPO}}(\theta) =  \mathbb{E}_{q, \{o_i\}} \bigg[ \frac{1}{\sum^{G}_i |o_i|} \sum_{i=1}^G \sum^{|o_i|}_{t=1} \min \Bigg( 
\frac{\pi_\theta(o_{i, t} \mid q)}{\pi_{\theta_{\text{old}}}(o_{i,t} \mid q)} A_{i}, 
\mathrm{clip} \left( \frac{\pi_\theta(o_{i,t} \mid q)}{\pi_{\theta_{\text{old}}}(o_{i,t} \mid q)}, 1 - \epsilon_{\text{low}}, 1 + \epsilon_{\text{high}} \right) A_i 
\Bigg) \bigg]
\end{aligned}
\end{equation}

where
\begin{equation}
A_{i} = \frac{R_i - \mathrm{mean}(\{ R_i \}_{i=1}^{G})}{\mathrm{std}(\{ R_i \}_{i=1}^{G})}
\end{equation}
is the  relative advantage computed from group-level reward signals. 
\( \epsilon_{\text{low}} \) and \( \epsilon_{\text{high}} \) denote the lower and upper clipping ratios, empirically set to 0.2 and 0.28, respectively.
Note that the loss is computed at the token level to better accommodate long-horizon reasoning patterns. In addition, we filtered the overlong samples to avoid reward noise in the training process.

\textbf{Rule-based Reward.} We use a rule-based reward. A response receives a reward of $+1$ if it passes all unit tests and $-1$ otherwise.

\textbf{Results.} Table~\ref{tab:livecodebench-compact} shows performance across difficulty levels. We observe consistent improvements in CodeContests\pl across all categories, especially on Easy and Medium tasks. This demonstrates the benefit of incorporating better test cases during training.

\section{Conclusion and Future Work}

In this paper, we propose an LLM-based Generator-Validator agent system capable of leveraging public problem data to construct high-quality test cases for competitive programming problems. This system facilitates the scaling up of high-quality code datasets. Using this agent system, we have developed CodeContests\pl\;by enhancing the original CodeContests dataset with better test cases. Experimental results demonstrate that the test cases in CodeContests\pl are of significantly higher quality than those in CodeContests, and that CodeContests\pl also exhibits substantial advantages in RL training.

According to our rough estimate, there are more than 100,000 programming problems with publicly available problem statements and ground truth solutions. Therefore, our proposed agent system will enable the full utilization of these data resources, thereby laying the data foundation for further enhancing the reasoning and coding capabilities of LLMs.

\section*{Acknowledgements}

We thank Jiaze Chen, Jinxin Chi, Zhicheng Liu, Siqian Chen, Jingjing Xu, Jinhua Zhu, Rui Long, Qi Liu, Li Han, Liang Xiang, as well as other colleagues at ByteDance, for their support for the CodeContests\pl\;project.

\clearpage

\bibliographystyle{plainnat}
\bibliography{main}

\begin{thebibliography}{18}
\providecommand{\natexlab}[1]{#1}
\providecommand{\url}[1]{\texttt{#1}}
\expandafter\ifx\csname urlstyle\endcsname\relax
  \providecommand{\doi}[1]{doi: #1}\else
  \providecommand{\doi}{doi: \begingroup \urlstyle{rm}\Url}\fi

\bibitem[Austin et~al.(2021)Austin, Odena, Nye, Bosma, Michalewski, Dohan, Jiang, Cai, Terry, Le, and Sutton]{mbpp}
Jacob Austin, Augustus Odena, Maxwell~I. Nye, Maarten Bosma, Henryk Michalewski, David Dohan, Ellen Jiang, Carrie~J. Cai, Michael Terry, Quoc~V. Le, and Charles Sutton.
\newblock Program synthesis with large language models.
\newblock \emph{CoRR}, abs/2108.07732, 2021.
\newblock URL \url{https://arxiv.org/abs/2108.07732}.

\bibitem[Chen et~al.(2025)Chen, Fan, Liu, Liu, Lin, Wang, Wang, Wei, Xu, Yuan, Yue, Yan, Yu, Zuo, Zhang, Zhu, An, Bai, Bao, Bin, Chen, Chen, Chen, Chen, Chen, Chen, Chen, Chen, Chen, Chen, Chen, Chen, Chi, Dai, Dai, Dai, Dou, Du, Du, Duan, Dun, Fan, Feng, Feng, Feng, Fu, Fu, Fu, Ge, Guo, Han, Han, Hao, Hao, He, He, He, Heng, Hong, Hou, Hu, Hu, Hu, Hua, Huang, Huang, Huang, Huang, Huang, Huang, Jia, Jia, Jia, Jiang, Jiang, Jiang, Jiang, Jiang, Jiao, Jin, Jin, Lai, Li, Li, Li, Li, Li, Wan, Wang, Li, Li, Li, Li, Li, Li, Li, Li, Liang, Liang, Lin, Lin, Lin, Liu, Liu, Liu, Liu, Liu, Liu, Liu, Liu, Liu, Liu, Liu, Liu, Liu, Liu, Liu, Liu, Long, Lou, Lou, Luo, Luo, Lv, Lv, Ma, Ma, Ma, Ma, Ma, Ma, Ma, Mao, Min, Nan, Ning, Ou, Pan, Pang, Peng, Peng, Qian, Qian, Qiao, Qu, Ren, Ren, Shan, Shen, Shen, Shen, Sheng, Shi, Shi, Shi, Cao, Song, Song, Su, Sun, Sun, Sun, Wan, Wang, Wang, Wang, Wang, Wang, Wang, Wang, Wang, Wang, Wang, Wang, Wang, Wang, Wang, Wang, Wang, Wang, Wei, Wei, Wei, Wei, Wu, Wu, Wu, Wu, Wu, Wu, Wu, Wu,
  Wu, Xi, Xia, Xian, Xiang, Xiang, Xiao, Xiao, Xiao, Xiao, Xin, Xin, Xiong, Xu, Xu, Xu, Xu, Xu, Xu, Xu, Xu, Yan, Yan, Yang, Yang, Yang, Yang, Yang, Yang, Yang, Yang, Yang, Yao, Yi, Yin, Yin, Ying, Yu, Yu, Yu, Yu, Yu, Yuan, Yuan, Zeng, Zhan, Zhang, Zhang, Zhang, Zhang, Zhang, Zhang, Zhang, Zhang, Zhang, Zhang, Zhang, Zhang, Zhang, Zhang, Zhang, Zhang, Zhang, Zheng, Zheng, Zheng, Zheng, Zheng, Zhi, Zhong, Zhong, Zhong, Zhong, Zhou, Zhou, Zhou, Zhu, Zhu, Zhu, and Zuo]{seedthinking}
Jiaze Chen, Tiantian Fan, Xin Liu, Lingjun Liu, Zhiqi Lin, Mingxuan Wang, Chengyi Wang, Xiangpeng Wei, Wenyuan Xu, Yufeng Yuan, Yu~Yue, Lin Yan, Qiying Yu, Xiaochen Zuo, Chi Zhang, Ruofei Zhu, Zhecheng An, Zhihao Bai, Yu~Bao, Xingyan Bin, Jiangjie Chen, Feng Chen, Hongmin Chen, Riwei Chen, Liangqiang Chen, Zixin Chen, Jinsong Chen, Siyan Chen, Kaiyuan Chen, Zhi Chen, Jin Chen, Jiecao Chen, Jinxin Chi, Weinan Dai, Ning Dai, Jiahui Dai, Shihan Dou, Yantao Du, Zhengyin Du, Jianhui Duan, Chen Dun, Ting-Han Fan, Jiazhan Feng, Junda Feng, Ziyuan Feng, Yuwei Fu, Wenqi Fu, Hanjie Fu, Hao Ge, Hongyi Guo, Mingji Han, Li~Han, Wenhao Hao, Xintong Hao, Qianyu He, Jerry He, Feng He, Wen Heng, Zehua Hong, Qi~Hou, Liang Hu, Shengding Hu, Nan Hu, Kai Hua, Qi~Huang, Ziyue Huang, Hongzhi Huang, Zihao Huang, Ting Huang, Wenhao Huang, Wei Jia, Bin Jia, Xiaoying Jia, Yuhua Jiang, Haobin Jiang, Ziheng Jiang, Kaihua Jiang, Chengquan Jiang, Jianpeng Jiao, Xiaoran Jin, Xing Jin, Xunhao Lai, Zheng Li, Xiang Li, Liyi Li, Hongkai Li,
  Zheng Li, Shengxian Wan, Ya~Wang, Yunshui Li, Chenggang Li, Niuniu Li, Siyu Li, Xi~Li, Xiao Li, Aoyan Li, Yuntao Li, Nianning Liang, Xinnian Liang, Haibin Lin, Weijian Lin, Ye~Lin, Zhicheng Liu, Guanlin Liu, Guanlin Liu, Chenxiao Liu, Yan Liu, Gaohong Liu, Juncai Liu, Chundian Liu, Deyi Liu, Kaibo Liu, Siyao Liu, Qi~Liu, Yongfei Liu, Kang Liu, Gan Liu, Boyi Liu, Rui Long, Weiqiang Lou, Chenwei Lou, Xiang Luo, Yao Luo, Caiping Lv, Heyang Lv, Bole Ma, Qianli Ma, Hongzhi Ma, Yiyuan Ma, Jin Ma, Wenchang Ma, Tingting Ma, Chen Mao, Qiyang Min, Zhe Nan, Guanghan Ning, Jinxiang Ou, Haojie Pan, Renming Pang, Yanghua Peng, Tao Peng, Lihua Qian, Lihua Qian, Mu~Qiao, Meng Qu, Cheng Ren, Hongbin Ren, Yong Shan, Wei Shen, Ke~Shen, Kai Shen, Guangming Sheng, Jinlong Shi, Wenlei Shi, Guang Shi, Shuai~Shuai Cao, Yuxin Song, Zuquan Song, Jing Su, Yifan Sun, Tao Sun, Zewei Sun, Borui Wan, Zihan Wang, Xiaohui Wang, Xi~Wang, Shuguang Wang, Jun Wang, Qinlong Wang, Chenyuan Wang, Shuai Wang, Zihan Wang, Changbao Wang, Jiaqiang
  Wang, Shihang Wang, Xuwu Wang, Zaiyuan Wang, Yuxuan Wang, Wenqi Wang, Taiqing Wang, Chengzhi Wei, Houmin Wei, Ziyun Wei, Shufa Wei, Zheng Wu, Yonghui Wu, Yangjun Wu, Bohong Wu, Shuang Wu, Jingqiao Wu, Ning Wu, Shuangzhi Wu, Jianmin Wu, Chenguang Xi, Fan Xia, Yuqiao Xian, Liang Xiang, Boren Xiang, Bowen Xiao, Zhen Xiao, Xia Xiao, Yongsheng Xiao, Chao Xin, Shulin Xin, Yuwen Xiong, Jingjing Xu, Ziwen Xu, Chenyin Xu, Jiayi Xu, Yifan Xu, Wei Xu, Yufei Xu, Shikun Xu, Shipeng Yan, Shen Yan, Qingping Yang, Xi~Yang, Tianhao Yang, Yuehang Yang, Yuan Yang, Ximing Yang, Zeyu Yang, Guang Yang, Yifan Yang, Xuesong Yao, Bairen Yi, Fan Yin, Jianian Yin, Ziqiang Ying, Xiangyu Yu, Hongli Yu, Song Yu, Menghan Yu, Huan Yu, Siyu Yuan, Jun Yuan, Yutao Zeng, Tianyang Zhan, Zheng Zhang, Yun Zhang, Mofan Zhang, Wang Zhang, Ru~Zhang, Zhi Zhang, Tianqi Zhang, Xinyi Zhang, Zhexi Zhang, Sijun Zhang, Wenqiang Zhang, Xiangxiang Zhang, Yongtao Zhang, Yuyu Zhang, Ge~Zhang, He~Zhang, Yue Zhang, Renjie Zheng, Ningxin Zheng, Zhuolin Zheng,
  Yaowei Zheng, Chen Zheng, Xiaoyun Zhi, Wanjun Zhong, Cheng Zhong, Zheng Zhong, Baoquan Zhong, Xun Zhou, Na~Zhou, Huan Zhou, Hang Zhu, Defa Zhu, Wenjia Zhu, and Lei Zuo.
\newblock Seed-thinking-v1.5: Advancing superb reasoning models with reinforcement learning, 2025.
\newblock URL \url{https://arxiv.org/abs/2504.13914}.

\bibitem[Chen et~al.(2021)Chen, Tworek, Jun, Yuan, de~Oliveira~Pinto, Kaplan, Edwards, Burda, Joseph, Brockman, Ray, Puri, Krueger, Petrov, Khlaaf, Sastry, Mishkin, Chan, Gray, Ryder, Pavlov, Power, Kaiser, Bavarian, Winter, Tillet, Such, Cummings, Plappert, Chantzis, Barnes, Herbert{-}Voss, Guss, Nichol, Paino, Tezak, Tang, Babuschkin, Balaji, Jain, Saunders, Hesse, Carr, Leike, Achiam, Misra, Morikawa, Radford, Knight, Brundage, Murati, Mayer, Welinder, McGrew, Amodei, McCandlish, Sutskever, and Zaremba]{humaneval}
Mark Chen, Jerry Tworek, Heewoo Jun, Qiming Yuan, Henrique~Pond{\'{e}} de~Oliveira~Pinto, Jared Kaplan, Harri Edwards, Yuri Burda, Nicholas Joseph, Greg Brockman, Alex Ray, Raul Puri, Gretchen Krueger, Michael Petrov, Heidy Khlaaf, Girish Sastry, Pamela Mishkin, Brooke Chan, Scott Gray, Nick Ryder, Mikhail Pavlov, Alethea Power, Lukasz Kaiser, Mohammad Bavarian, Clemens Winter, Philippe Tillet, Felipe~Petroski Such, Dave Cummings, Matthias Plappert, Fotios Chantzis, Elizabeth Barnes, Ariel Herbert{-}Voss, William~Hebgen Guss, Alex Nichol, Alex Paino, Nikolas Tezak, Jie Tang, Igor Babuschkin, Suchir Balaji, Shantanu Jain, William Saunders, Christopher Hesse, Andrew~N. Carr, Jan Leike, Joshua Achiam, Vedant Misra, Evan Morikawa, Alec Radford, Matthew Knight, Miles Brundage, Mira Murati, Katie Mayer, Peter Welinder, Bob McGrew, Dario Amodei, Sam McCandlish, Ilya Sutskever, and Wojciech Zaremba.
\newblock Evaluating large language models trained on code.
\newblock \emph{CoRR}, abs/2107.03374, 2021.
\newblock URL \url{https://arxiv.org/abs/2107.03374}.

\bibitem[Cheng et~al.(2024)Cheng, Chen, Chen, Chen, Chen, Chen, Chen, Geng, Li, Li, Li, Li, Liu, Liu, Liu, Liu, Liu, Liu, Liu, Liu, Liu, Long, Mai, Ning, Peng, Shen, Su, Su, Sun, Sun, Tao, Wang, Wang, Wang, Wang, Wang, Xia, Xiang, Xiao, Xiao, Xi, Xin, Xu, Xu, Yang, Yang, Yang, Yuan, Zhang, Zhang, Zhang, Zheng, Zhu, and Zhu]{sandbox}
Yao Cheng, Jianfeng Chen, Jie Chen, Li~Chen, Liyu Chen, Wentao Chen, Zhengyu Chen, Shijie Geng, Aoyan Li, Bo~Li, Bowen Li, Linyi Li, Boyi Liu, Jerry Liu, Kaibo Liu, Qi~Liu, Shukai Liu, Siyao Liu, Tianyi Liu, Tingkai Liu, Yongfei Liu, Rui Long, Jing Mai, Guanghan Ning, Z.~Y. Peng, Kai Shen, Jiahao Su, Jing Su, Tao Sun, Yifan Sun, Yunzhe Tao, Guoyin Wang, Siwei Wang, Xuwu Wang, Yite Wang, Zihan Wang, Jinxiang Xia, Liang Xiang, Xia Xiao, Yongsheng Xiao, Chenguang Xi, Shulin Xin, Jingjing Xu, Shikun Xu, Hongxia Yang, Jack Yang, Yingxiang Yang, Jianbo Yuan, Jun Zhang, Yufeng Zhang, Yuyu Zhang, Shen Zheng, He~Zhu, and Ming Zhu.
\newblock {FullStack Bench}: Evaluating {LLMs} as full stack coders, 2024.
\newblock URL \url{https://arxiv.org/abs/2412.00535}.

\bibitem[El-Kishky et~al.(2025)El-Kishky, Wei, Saraiva, Minaiev, Selsam, Dohan, Song, Lightman, Clavera, Pachocki, Tworek, Kuhn, Kaiser, Chen, Schwarzer, Rohaninejad, McAleese, o3~contributors, Mürk, Garg, Shu, Sidor, Kosaraju, and Zhou]{o1}
Ahmed El-Kishky, Alexander Wei, Andre Saraiva, Borys Minaiev, Daniel Selsam, David Dohan, Francis Song, Hunter Lightman, Ignasi Clavera, Jakub Pachocki, Jerry Tworek, Lorenz Kuhn, Lukasz Kaiser, Mark Chen, Max Schwarzer, Mostafa Rohaninejad, Nat McAleese, o3~contributors, Oleg Mürk, Rhythm Garg, Rui Shu, Szymon Sidor, Vineet Kosaraju, and Wenda Zhou.
\newblock Competitive programming with large reasoning models, 2025.
\newblock URL \url{https://arxiv.org/abs/2502.06807}.

\bibitem[Guo et~al.(2025)Guo, Yang, Zhang, Song, Zhang, Xu, Zhu, Ma, Wang, Bi, Zhang, Yu, Wu, Wu, Gou, Shao, Li, Gao, Liu, Xue, Wang, Wu, Feng, Lu, Zhao, Deng, Zhang, Ruan, Dai, Chen, Ji, Li, Lin, Dai, Luo, Hao, Chen, Li, Zhang, Bao, Xu, Wang, Ding, Xin, Gao, Qu, Li, Guo, Li, Wang, Chen, Yuan, Qiu, Li, Cai, Ni, Liang, Chen, Dong, Hu, Gao, Guan, Huang, Yu, Wang, Zhang, Zhao, Wang, Zhang, Xu, Xia, Zhang, Zhang, Tang, Li, Wang, Li, Tian, Huang, Zhang, Wang, Chen, Du, Ge, Zhang, Pan, Wang, Chen, Jin, Chen, Lu, Zhou, Chen, Ye, Wang, Yu, Zhou, Pan, Li, Zhou, Wu, Ye, Yun, Pei, Sun, Wang, Zeng, Zhao, Liu, Liang, Gao, Yu, Zhang, Xiao, An, Liu, Wang, Chen, Nie, Cheng, Liu, Xie, Liu, Yang, Li, Su, Lin, Li, Jin, Shen, Chen, Sun, Wang, Song, Zhou, Wang, Shan, Li, Wang, Wei, Zhang, Xu, Li, Zhao, Sun, Wang, Yu, Zhang, Shi, Xiong, He, Piao, Wang, Tan, Ma, Liu, Guo, Ou, Wang, Gong, Zou, He, Xiong, Luo, You, Liu, Zhou, Zhu, Xu, Huang, Li, Zheng, Zhu, Ma, Tang, Zha, Yan, Ren, Ren, Sha, Fu, Xu, Xie, Zhang, Hao, Ma, Yan, Wu, Gu,
  Zhu, Liu, Li, Xie, Song, Pan, Huang, Xu, Zhang, and Zhang]{deepseekr1}
Daya Guo, Dejian Yang, Haowei Zhang, Junxiao Song, Ruoyu Zhang, Runxin Xu, Qihao Zhu, Shirong Ma, Peiyi Wang, Xiao Bi, Xiaokang Zhang, Xingkai Yu, Yu~Wu, Z.~F. Wu, Zhibin Gou, Zhihong Shao, Zhuoshu Li, Ziyi Gao, Aixin Liu, Bing Xue, Bingxuan Wang, Bochao Wu, Bei Feng, Chengda Lu, Chenggang Zhao, Chengqi Deng, Chenyu Zhang, Chong Ruan, Damai Dai, Deli Chen, Dongjie Ji, Erhang Li, Fangyun Lin, Fucong Dai, Fuli Luo, Guangbo Hao, Guanting Chen, Guowei Li, H.~Zhang, Han Bao, Hanwei Xu, Haocheng Wang, Honghui Ding, Huajian Xin, Huazuo Gao, Hui Qu, Hui Li, Jianzhong Guo, Jiashi Li, Jiawei Wang, Jingchang Chen, Jingyang Yuan, Junjie Qiu, Junlong Li, J.~L. Cai, Jiaqi Ni, Jian Liang, Jin Chen, Kai Dong, Kai Hu, Kaige Gao, Kang Guan, Kexin Huang, Kuai Yu, Lean Wang, Lecong Zhang, Liang Zhao, Litong Wang, Liyue Zhang, Lei Xu, Leyi Xia, Mingchuan Zhang, Minghua Zhang, Minghui Tang, Meng Li, Miaojun Wang, Mingming Li, Ning Tian, Panpan Huang, Peng Zhang, Qiancheng Wang, Qinyu Chen, Qiushi Du, Ruiqi Ge, Ruisong Zhang,
  Ruizhe Pan, Runji Wang, R.~J. Chen, R.~L. Jin, Ruyi Chen, Shanghao Lu, Shangyan Zhou, Shanhuang Chen, Shengfeng Ye, Shiyu Wang, Shuiping Yu, Shunfeng Zhou, Shuting Pan, S.~S. Li, Shuang Zhou, Shaoqing Wu, Shengfeng Ye, Tao Yun, Tian Pei, Tianyu Sun, T.~Wang, Wangding Zeng, Wanjia Zhao, Wen Liu, Wenfeng Liang, Wenjun Gao, Wenqin Yu, Wentao Zhang, W.~L. Xiao, Wei An, Xiaodong Liu, Xiaohan Wang, Xiaokang Chen, Xiaotao Nie, Xin Cheng, Xin Liu, Xin Xie, Xingchao Liu, Xinyu Yang, Xinyuan Li, Xuecheng Su, Xuheng Lin, X.~Q. Li, Xiangyue Jin, Xiaojin Shen, Xiaosha Chen, Xiaowen Sun, Xiaoxiang Wang, Xinnan Song, Xinyi Zhou, Xianzu Wang, Xinxia Shan, Y.~K. Li, Y.~Q. Wang, Y.~X. Wei, Yang Zhang, Yanhong Xu, Yao Li, Yao Zhao, Yaofeng Sun, Yaohui Wang, Yi~Yu, Yichao Zhang, Yifan Shi, Yiliang Xiong, Ying He, Yishi Piao, Yisong Wang, Yixuan Tan, Yiyang Ma, Yiyuan Liu, Yongqiang Guo, Yuan Ou, Yuduan Wang, Yue Gong, Yuheng Zou, Yujia He, Yunfan Xiong, Yuxiang Luo, Yuxiang You, Yuxuan Liu, Yuyang Zhou, Y.~X. Zhu, Yanhong Xu,
  Yanping Huang, Yaohui Li, Yi~Zheng, Yuchen Zhu, Yunxian Ma, Ying Tang, Yukun Zha, Yuting Yan, Z.~Z. Ren, Zehui Ren, Zhangli Sha, Zhe Fu, Zhean Xu, Zhenda Xie, Zhengyan Zhang, Zhewen Hao, Zhicheng Ma, Zhigang Yan, Zhiyu Wu, Zihui Gu, Zijia Zhu, Zijun Liu, Zilin Li, Ziwei Xie, Ziyang Song, Zizheng Pan, Zhen Huang, Zhipeng Xu, Zhongyu Zhang, and Zhen Zhang.
\newblock {DeepSeek-R1}: Incentivizing reasoning capability in {LLMs} via reinforcement learning, 2025.
\newblock URL \url{https://arxiv.org/abs/2501.12948}.

\bibitem[Hendrycks et~al.(2021)Hendrycks, Basart, Kadavath, Mazeika, Arora, Guo, Burns, Puranik, He, Song, and Steinhardt]{apps}
Dan Hendrycks, Steven Basart, Saurav Kadavath, Mantas Mazeika, Akul Arora, Ethan Guo, Collin Burns, Samir Puranik, Horace He, Dawn Song, and Jacob Steinhardt.
\newblock Measuring coding challenge competence with {APPS}.
\newblock In Joaquin Vanschoren and Sai{-}Kit Yeung, editors, \emph{Proceedings of the Neural Information Processing Systems Track on Datasets and Benchmarks 1, NeurIPS Datasets and Benchmarks 2021, December 2021, virtual}, 2021.
\newblock URL \url{https://datasets-benchmarks-proceedings.neurips.cc/paper/2021/hash/c24cd76e1ce41366a4bbe8a49b02a028-Abstract-round2.html}.

\bibitem[Jain et~al.(2025)Jain, Han, Gu, Li, Yan, Zhang, Wang, Solar{-}Lezama, Sen, and Stoica]{livecodebench}
Naman Jain, King Han, Alex Gu, Wen{-}Ding Li, Fanjia Yan, Tianjun Zhang, Sida Wang, Armando Solar{-}Lezama, Koushik Sen, and Ion Stoica.
\newblock {LiveCodeBench}: Holistic and contamination free evaluation of large language models for code.
\newblock In \emph{The Twelfth International Conference on Learning Representations, {ICLR} 2025, Singapore, April 24-28, 2025}. OpenReview.net, 2025.
\newblock URL \url{https://openreview.net/forum?id=chfJJYC3iL}.

\bibitem[Jiang(2016)]{tlb}
Yichuan Jiang.
\newblock A survey of task allocation and load balancing in distributed systems.
\newblock \emph{{IEEE} Trans. Parallel Distributed Syst.}, 27\penalty0 (2):\penalty0 585--599, 2016.
\newblock \doi{10.1109/TPDS.2015.2407900}.
\newblock URL \url{https://doi.org/10.1109/TPDS.2015.2407900}.

\bibitem[Li et~al.(2023)Li, Fu, Zhang, Huang, Sun, Lyu, Liu, Jin, and Li]{taco}
Rongao Li, Jie Fu, Bo{-}Wen Zhang, Tao Huang, Zhihong Sun, Chen Lyu, Guang Liu, Zhi Jin, and Ge~Li.
\newblock {TACO:} topics in algorithmic code generation dataset.
\newblock \emph{CoRR}, abs/2312.14852, 2023.
\newblock \doi{10.48550/ARXIV.2312.14852}.
\newblock URL \url{https://doi.org/10.48550/arXiv.2312.14852}.

\bibitem[Li et~al.(2022)Li, Choi, Chung, Kushman, Schrittwieser, Leblond, Eccles, Keeling, Gimeno, Lago, Hubert, Choy, de~Masson~d’Autume, Babuschkin, Chen, Huang, Welbl, Gowal, Cherepanov, Molloy, Mankowitz, Robson, Kohli, de~Freitas, Kavukcuoglu, and Vinyals]{alphacode}
Yujia Li, David Choi, Junyoung Chung, Nate Kushman, Julian Schrittwieser, Rémi Leblond, Tom Eccles, James Keeling, Felix Gimeno, Agustin~Dal Lago, Thomas Hubert, Peter Choy, Cyprien de~Masson~d’Autume, Igor Babuschkin, Xinyun Chen, Po-Sen Huang, Johannes Welbl, Sven Gowal, Alexey Cherepanov, James Molloy, Daniel~J. Mankowitz, Esme~Sutherland Robson, Pushmeet Kohli, Nando de~Freitas, Koray Kavukcuoglu, and Oriol Vinyals.
\newblock Competition-level code generation with {AlphaCode}.
\newblock \emph{Science}, 378\penalty0 (6624):\penalty0 1092--1097, 2022.
\newblock \doi{10.1126/science.abq1158}.
\newblock URL \url{https://www.science.org/doi/abs/10.1126/science.abq1158}.

\bibitem[Liu et~al.(2023)Liu, Xia, Wang, and Zhang]{evalplus}
Jiawei Liu, Chunqiu~Steven Xia, Yuyao Wang, and Lingming Zhang.
\newblock Is your code generated by {ChatGPT} really correct? {Rigorous} evaluation of large language models for code generation.
\newblock In Alice Oh, Tristan Naumann, Amir Globerson, Kate Saenko, Moritz Hardt, and Sergey Levine, editors, \emph{Advances in Neural Information Processing Systems 36: Annual Conference on Neural Information Processing Systems 2023, NeurIPS 2023, New Orleans, LA, USA, December 10 - 16, 2023}, 2023.
\newblock URL \url{http://papers.nips.cc/paper\_files/paper/2023/hash/43e9d647ccd3e4b7b5baab53f0368686-Abstract-Conference.html}.

\bibitem[Mirzayanov(2005)]{testlib}
Mike Mirzayanov.
\newblock Testlib, 2005.
\newblock URL \url{https://github.com/MikeMirzayanov/testlib}.

\bibitem[Mirzayanov et~al.(2020)Mirzayanov, Pavlova, MAVRIN, Melnikov, Plotnikov, Parfenov, and Stankevich]{cf}
Mike Mirzayanov, Oksana Pavlova, Pavel MAVRIN, Roman Melnikov, Andrew Plotnikov, Vladimir Parfenov, and Andrew Stankevich.
\newblock Codeforces as an educational platform for learning programming in digitalization.
\newblock \emph{Olympiads in Informatics}, 14\penalty0 (133-142):\penalty0 14, 2020.

\bibitem[Shi et~al.(2024)Shi, Tang, Narasimhan, and Yao]{usaco}
Quan Shi, Michael Tang, Karthik Narasimhan, and Shunyu Yao.
\newblock Can language models solve olympiad programming?, 2024.
\newblock URL \url{https://arxiv.org/abs/2404.10952}.

\bibitem[Shojaee et~al.(2023)Shojaee, Jain, Tipirneni, and Reddy]{shojaee2023ppocoder}
Parshin Shojaee, Aneesh Jain, Sindhu Tipirneni, and Chandan~K. Reddy.
\newblock Execution-based code generation using deep reinforcement learning.
\newblock \emph{Trans. Mach. Learn. Res.}, 2023, 2023.
\newblock URL \url{https://openreview.net/forum?id=0XBuaxqEcG}.

\bibitem[Yang et~al.(2025)Yang, Yang, Zhang, Hui, Zheng, Yu, Li, Liu, Huang, Wei, Lin, Yang, Tu, Zhang, Yang, Yang, Zhou, Lin, Dang, Lu, Bao, Yang, Yu, Li, Xue, Zhang, Zhu, Men, Lin, Li, Tang, Xia, Ren, Ren, Fan, Su, Zhang, Wan, Liu, Cui, Zhang, and Qiu]{qwen2024technical}
An~Yang, Baosong Yang, Beichen Zhang, Binyuan Hui, Bo~Zheng, Bowen Yu, Chengyuan Li, Dayiheng Liu, Fei Huang, Haoran Wei, Huan Lin, Jian Yang, Jianhong Tu, Jianwei Zhang, Jianxin Yang, Jiaxi Yang, Jingren Zhou, Junyang Lin, Kai Dang, Keming Lu, Keqin Bao, Kexin Yang, Le~Yu, Mei Li, Mingfeng Xue, Pei Zhang, Qin Zhu, Rui Men, Runji Lin, Tianhao Li, Tianyi Tang, Tingyu Xia, Xingzhang Ren, Xuancheng Ren, Yang Fan, Yang Su, Yichang Zhang, Yu~Wan, Yuqiong Liu, Zeyu Cui, Zhenru Zhang, and Zihan Qiu.
\newblock Qwen2.5 technical report, 2025.
\newblock URL \url{https://arxiv.org/abs/2412.15115}.

\bibitem[Yu et~al.(2025)Yu, Zhang, Zhu, Yuan, Zuo, Yue, Dai, Fan, Liu, Liu, Liu, Lin, Lin, Ma, Sheng, Tong, Zhang, Zhang, Zhang, Zhu, Zhu, Chen, Chen, Wang, Yu, Song, Wei, Zhou, Liu, Ma, Zhang, Yan, Qiao, Wu, and Wang]{yu2025dapo}
Qiying Yu, Zheng Zhang, Ruofei Zhu, Yufeng Yuan, Xiaochen Zuo, Yu~Yue, Weinan Dai, Tiantian Fan, Gaohong Liu, Lingjun Liu, Xin Liu, Haibin Lin, Zhiqi Lin, Bole Ma, Guangming Sheng, Yuxuan Tong, Chi Zhang, Mofan Zhang, Wang Zhang, Hang Zhu, Jinhua Zhu, Jiaze Chen, Jiangjie Chen, Chengyi Wang, Hongli Yu, Yuxuan Song, Xiangpeng Wei, Hao Zhou, Jingjing Liu, Wei-Ying Ma, Ya-Qin Zhang, Lin Yan, Mu~Qiao, Yonghui Wu, and Mingxuan Wang.
\newblock {DAPO}: An open-source {LLM} reinforcement learning system at scale, 2025.
\newblock URL \url{https://arxiv.org/abs/2503.14476}.

\end{thebibliography}

\clearpage

\beginappendix

\section{Terms and Definitions}

The terms used in this paper and their definitions are summarized as follows.

\textbf{Submission.} In programming competitions, a submission is a program submitted by a contestant. This submission is evaluated by a judging system, resulting in a verdict such as Accepted, Wrong Answer, Time Limit Exceeded, Runtime Error, or Compile Error, among others.

\textbf{Test case, test data.} In competitive programming, a test case is used to check whether the participant's submission is correct. It usually consists of test input and test output.

\textbf{Test input, input data.} The input data of a test case will be fed into the contestant's program. The output obtained will then be compared with the ground truth data.
        
\textbf{Test output, output data, reference answer.} The output data of a test case is the correct answer corresponding to the input data. For problems with multiple correct solutions, the output data typically represents one of the possible correct answers.

\textbf{Validator, input validator, validator program.} A validator is a program used to check if input data is correct.

\textbf{Generator, input generator, generator program.} Because some input data is very large or complex and cannot be handcrafted, a generator is used to produce it. A generator is a program that creates test input.

\textbf{Checker, output checker, special judge.} A checker is a program used to determine if a contestant's output is correct. Usually, it simply checks if the contestant's output matches the test output. For problems with multiple solutions, it may contain more complex logic.

\begin{figure}
    \centering
    \includegraphics[width=1\linewidth]{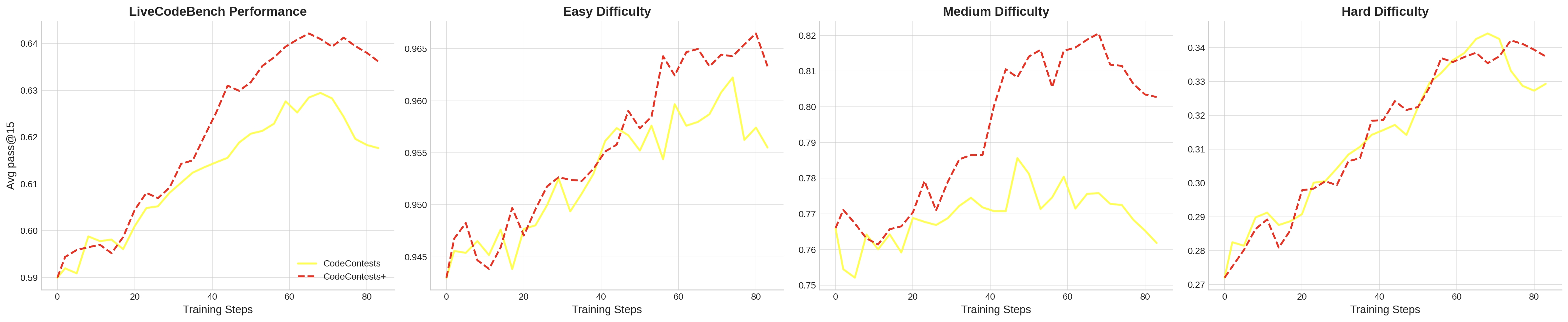}
    \caption{Evaluation results during RL training process.}
    \label{fig:rl_process}
\end{figure}

\section{Checker}
\label{sec:checker}

\textbf{Checker Program.} Previous research has largely employed a simple character-by-character comparison to determine if a program's output is correct. However, this approach does not apply to problems with multiple solutions. Many problems in programming competitions have multiple valid solutions, meaning the same input can correspond to multiple correct outputs. For example, approximately 1/4 of the problems on Codeforces are multi-solution problems. These types of problems require special checking logic. A Checker is a program used to check if a program's output is correct; its inputs include the input data, program output, and the reference answer, and it determines if the program output is correct. For problems without multiple solutions, a checker program may still be necessary. For example, problems with floating-point outputs need to compute relative or absolute errors to judge the correctness of the answer. Even for the most general problems, a checker program may need simple logic to ignore extra spaces or line breaks from the program's output. Therefore, we believe providing a custom checker program for every problem is necessary. A checker example is presented in Appendix \ref{sec:demogenerator}.

Additionally, the checker can provide richer error feedback than a simple correct/incorrect binary label. For example, the checker can specify the exact location of the error in the output, the differences between the correct answer and the program's output, and even the reason for the error. This information can be utilized in future research, for example, to help Code LLMs with reflection.

\textbf{Agent Workflow.} First, the checker agent is provided with a problem statement. Next, the agent is required to carefully read the problem statement and determine if the problem has multiple valid solutions. Multi-solution problems typically have explicit hints, such as "If multiple feasible solutions exist, output any one of them." If the agent determines that the problem does not have multiple solutions, the agent then needs to select one of the built-in checkers based on the problem's output format. These built-in checkers include an integer checker, a floating-point checker, a yes/no checker, and a general full-text comparison checker, among others. A complete list of built-in checkers is presented in Table \ref{tab:checkers}. If the agent determines that the problem has multiple solutions, then it needs to implement custom checking logic based on the problem requirements and output a checker program.

\textbf{Supervision.} Based on our observations, the main errors made by the checker agent are concentrated in multi-solution problems where the checker is more difficult to implement, and more often, these errors involve incorrectly checker correct answers as incorrect. We use a relatively simple supervision method where we input the problem's sample input and sample output into the checker (with the sample output serving as both the program's output and the reference answer). If the checker fails, it indicates an error in the checker's implementation. In such cases, the sample input, sample output, and the checker's output are all fed back to the checker agent for reflection and correction. Even with this, there are still some problems where the agent is unable to correctly implement the checker. These case studies will be presented in the Section \ref{sec:evaluation}.

To ease the burden on the agent, for problems without multiple valid solutions, we provide the agent with a selection of pre-written checkers, eliminating the need for the agent to write one itself. These built-in checkers are presented in Table \ref{tab:checkers}.

\begin{table*}
    \caption{Built-in checkers and their uses.}
    \vspace{1.5pt}
    \footnotesize
    \centering
    \begin{tabular}{ll}
    \toprule
       Name  & Use\\
       \midrule
        \texttt{ncmp.cc} & Compare ordered sequences of signed 64-bit integer numbers.\\
        \texttt{rcmp4.cc} & Compare two sequences of floating point numbers, with max absolute or relative error = $10^{-4}$.\\
        \texttt{rcmp6.cc} &Compare two sequences of floating point numbers, with max absolute or relative error = $10^{-6}$. \\
        \texttt{rcmp9.cc} & Compare two sequences of floating point numbers, with max absolute or relative error = $10^{-9}$.\\
        \texttt{wcmp.cc} & Compare sequences of tokens. Invisible characters are regarded as separators and are not compared.\\
        \texttt{hcmp.cc} & Compare two signed huge integers.\\
        \texttt{nyesno.cc} & Compare multiple \texttt{YES} or \texttt{NO} (case insensitive).\\
        \texttt{fcmp.cc} & Full-text comparison. Whitespaces, tabs, and linebreaks are also strictly compared.\\
         \bottomrule
    \end{tabular}
    
    \label{tab:checkers}
\end{table*}

\section{Evaluation as a Service (EaaS): A Cloud Architecture for Large-Scale Code Evaluation}
\label{sec:cloud}

Training Code LLMs with RL requires sampling a large amount of solutions and evaluating them to obtain rewards for model training. Therefore, code evaluation has become a bottleneck affecting training efficiency, with its computational overhead being comparable to the LLM’s parameter updates. To address this, we implemented a cloud service for large-scale code evaluation. 

This cloud service runs on a cluster of 25,000 CPU cores, where 8,000 cores comprise 2,000 \texttt{4c16g} judging pods, and 17,000 cores comprise 8,500 \texttt{2c4g} execution pods. Each judging pod is responsible for evaluating an entire code request, while each execution pod \cite{sandbox} is responsible for running a code with a single test input and returning the output. The judging pod passes the code and input files to the execution pod; after execution, the execution pod returns the output to the judging pod, which runs a checker to determine the correctness of the output. Execution pods have a concurrency of 1 and exclusively use two cores of the host machine, ensuring both isolation and consistent performance. Judging pods, on the other hand, have higher concurrency and do not need to run with exclusive cores. Execution pods are scheduled using load balancing \cite{tlb} while judging pods are scheduled using a message queue. All test cases are preprocessed and synchronized across judging pods via a network file system, which eliminates the time overhead of compiling and running generators. Execution pods are equipped with runtime environments for multiple languages, including C, C++, Java, Python, Rust, Go, C\#, and PHP, among others.

\section{Illustrations, Demos and Cases}
\label{sec:cases}

\subsection{Demo: Validator}
\label{sec:demovalidator}

\begin{figure}
    \centering
    \includegraphics[width=0.5\linewidth]{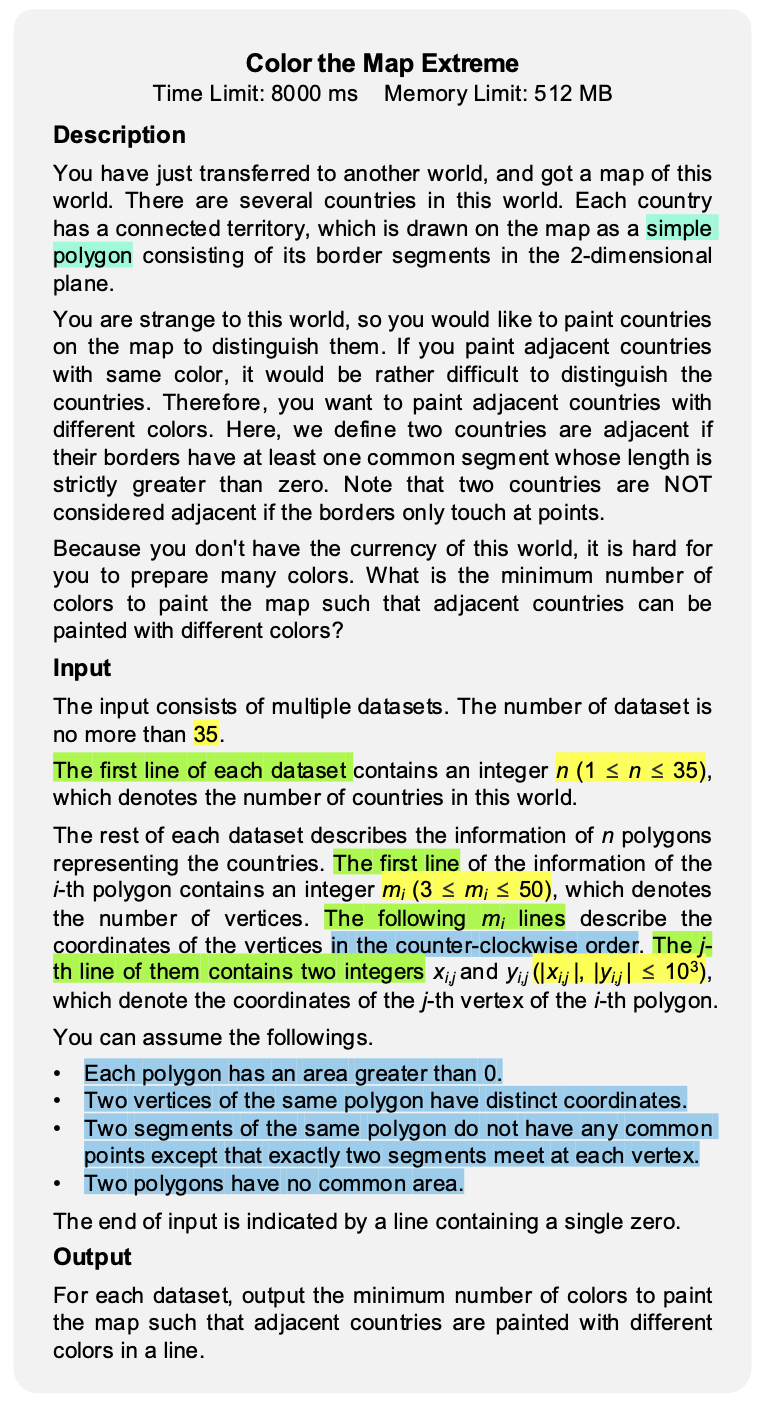}
    \caption{An example problem \textit{Color the Map Extreme}.}
    \label{fig:example_problem}
\end{figure}

The agent-written validator of the problem \textit{Color the Map Extreme}, which is presented in Fig. \ref{fig:example_problem}, is as follow:

\begin{lstlisting}[language=c++]
#include "testlib.h"
#include <bits/stdc++.h>
using namespace std;

struct Point {
    long long x, y;
};

static long long cross(const Point &A, const Point &B, const Point &C) {
    // Returns the cross product (B-A) x (C-A).
    // Positive if A->B->C is counter-clockwise turn,
    // negative if clockwise, 0 if collinear.
    return (B.x - A.x) * (C.y - A.y)
         - (B.y - A.y) * (C.x - A.x);
}

// Check if two line segments (p1,p2) and (p3,p4) intersect in a way
// that produces a 2D area overlap (i.e., truly crossing).
// We *allow* them to share endpoints or overlap on boundary with no interior crossing.
bool segmentsProperlyIntersect(const Point &p1, const Point &p2,
                               const Point &p3, const Point &p4)
{
    long long c1 = cross(p1, p2, p3);
    long long c2 = cross(p1, p2, p4);
    long long c3 = cross(p3, p4, p1);
    long long c4 = cross(p3, p4, p2);

    // If the two segments share more than a point on the boundary in a line overlap,
    // that is still not an "area" intersection. We consider a "proper" intersection
    // if each segment straddles the line extended by the other segment
    // (typical computational geometry check for crossing).
    if ((c1 == 0 && c2 == 0 && c3 == 0 && c4 == 0)) {
        // Collinear or overlapping segments on same line:
        // We check bounding boxes for overlap. That's still not 2D overlap (area).
        // So do not treat as an area intersection.
        return false;
    }

    // If there's a genuine crossing, each cross product pair has different sign
    // (i.e. c1 and c2 differ in sign, c3 and c4 differ in sign).
    // That means a proper crossing that yields an intersection in the interior.
    bool s1 = (c1 < 0 && c2 > 0) || (c1 > 0 && c2 < 0);
    bool s2 = (c3 < 0 && c4 > 0) || (c3 > 0 && c4 < 0);
    return (s1 && s2);
}

// Checks if point P is strictly inside the polygon poly (not on boundary).
// Standard ray casting or winding approach. Here we'll use winding number
// or cross approach. We exclude boundary checks, so if it is on boundary
// we'll consider that "not strictly inside."
bool pointInPolygonStrict(const vector<Point> &poly, const Point &P) {
    // A typical approach is to check the sum of angles or the winding number.
    // We'll do a ray-casting approach that ignores boundary if crossing hits a vertex.
    // Because coordinates can be negative and up to 1000 in absolute value, we'll do
    // the standard approach carefully with integer checks.
    int cnt = 0;
    for (int i = 0; i < (int)poly.size(); i++) {
        Point A = poly[i];
        Point B = poly[(i + 1) % poly.size()];
        // Check if the ray from P horizontally to the right intersects segment AB.
        // We'll skip boundary checks in a way that if P is exactly on an edge, we return false.
        if ((A.y > P.y) != (B.y > P.y)) {
            // Possible crossing in y
            long double x_cross = (long double)(B.x - A.x) * (long double)(P.y - A.y)
                                / (long double)(B.y - A.y) + A.x;
            if (x_cross > P.x) {
                cnt++;
            }
        }
        // Also check if P is exactly on segment AB. If so, it's not strictly inside.
        long long c = cross(A, B, P);
        if (c == 0) {
            // Check if P is within bounding box of A and B
            if (min(A.x, B.x) <= P.x && P.x <= max(A.x, B.x) &&
                min(A.y, B.y) <= P.y && P.y <= max(A.y, B.y)) {
                // On boundary
                return false;
            }
        }
    }
    // If odd, inside; if even, outside. 
    // Because we skip boundary as "strict inside," we only count if definitely inside.
    return (cnt % 2 == 1);
}

// Check if polygon is simple (no self intersections) 
// and has strictly positive area (CCW is required).
void checkSimplePolygon(const vector<Point> &poly) {
    int m = (int)poly.size();
    // 1) Check no two vertices coincide.
    // 2) Check edges do not intersect except at adjacent vertices.
    // 3) Check area > 0 and orientation is CCW.

    // Distinct vertices:
    set<pair<long long, long long>> st;
    for (auto &pt : poly) {
        auto p = make_pair(pt.x, pt.y);
        ensuref(!st.count(p), "Polygon has duplicate vertices.");
        st.insert(p);
    }

    // Check area (signed area for CCW).
    // Standard shoelace formula:
    long long area2 = 0; // 2 * area, signed
    for (int i = 0; i < m; i++) {
        int j = (i + 1) % m;
        area2 += (poly[i].x * poly[j].y - poly[j].x * poly[i].y);
    }
    ensuref(area2 > 0, "Polygon must have positive area and CCW order (area2=%lld)", area2);

    // Check self-intersections:
    // We'll compare every pair of edges (i, i+1) with (j, j+1), skipping adjacency.
    for (int i = 0; i < m; i++) {
        Point p1 = poly[i];
        Point p2 = poly[(i + 1) % m];
        for (int j = i + 1; j < m; j++) {
            // edges (i->i+1) and (j->j+1) share endpoints if j == i+1 or i == j+1
            // (mod m). We'll skip adjacency or the same edge.
            if (abs(i - j) <= 1 || (i == 0 && j == m - 1) || (j == 0 && i == m - 1)) {
                continue; // adjacent edges or the same edge
            }
            Point p3 = poly[j];
            Point p4 = poly[(j + 1) % m];
            // If a proper intersection occurs, it's invalid:
            if (segmentsProperlyIntersect(p1, p2, p3, p4)) {
                ensuref(false, "Polygon edges intersect in their interiors.");
            }
        }
    }
}

// Check if polygons A and B share any area. 
// Return true if they do share area > 0, false if they only meet at boundary or do not meet at all.
bool polygonsShareArea(const vector<Point> &A, const vector<Point> &B) {
    // We'll do:
    // 1) Check if any vertex of A is strictly inside B => area overlap
    // 2) Check if any vertex of B is strictly inside A => area overlap
    // 3) Check if edges properly intersect (cross) => area overlap
    // If none of these => no area overlap. 
    // Overlapping in a line or point only is fine (no area).
    for (auto &p : A) {
        if (pointInPolygonStrict(B, p)) {
            return true;
        }
    }
    for (auto &p : B) {
        if (pointInPolygonStrict(A, p)) {
            return true;
        }
    }
    // Check edges pairwise for a crossing intersection
    for (int i = 0; i < (int)A.size(); i++) {
        Point p1 = A[i];
        Point p2 = A[(i + 1) % A.size()];
        for (int j = 0; j < (int)B.size(); j++) {
            Point p3 = B[j];
            Point p4 = B[(j + 1) % B.size()];
            if (segmentsProperlyIntersect(p1, p2, p3, p4)) {
                return true;
            }
        }
    }
    return false;
}

int main(int argc, char* argv[]) {
    registerValidation(argc, argv);

    while (true) {
        int n = inf.readInt(0, 35); // n can be zero or up to 35
        inf.readEoln();
        if (n == 0) {
            // End of input
            break;
        }

        // Read n polygons
        vector<vector<Point>> polygons(n);
        for (int i = 0; i < n; i++) {
            int m = inf.readInt(3, 50);
            inf.readEoln();
            vector<Point> poly(m);
            for (int j = 0; j < m; j++) {
                long long x = inf.readLong(-1000, 1000);
                inf.readSpace();
                long long y = inf.readLong(-1000, 1000);
                inf.readEoln();
                poly[j] = {x, y};
            }
            // Check that this polygon is valid (simple, area>0, CCW).
            checkSimplePolygon(poly);
            polygons[i] = poly;
        }

        // Check that no two polygons share area.
        // (They may only meet at boundary or be disjoint.)
        for (int i = 0; i < n; i++) {
            for (int j = i + 1; j < n; j++) {
                bool hasAreaOverlap = polygonsShareArea(polygons[i], polygons[j]);
                ensuref(!hasAreaOverlap, "Two polygons share area (overlap).");
            }
        }
    }

    inf.readEof();
    return 0;
}
\end{lstlisting}

\subsection{Demo: Generator and Checker}
\label{sec:demogenerator}
Problem: \href{https://codeforces.com/problemset/problem/152/E}{https://codeforces.com/problemset/problem/152/E}

The generator of this problem is as follows:

\begin{lstlisting}[language=c++]
#include "testlib.h"
#include <bits/stdc++.h>
using namespace std;

/*
  This generator produces a single test case for the "Vasya's Garden" problem.
  It reads four arguments: n, m, k, and type (default = "random").

  Constraints to satisfy:
    1 <= n, m <= 100
    n*m <= 200
    1 <= k <= min(n*m, 7)
    1 <= a[i][j] <= 1000
    All k important squares must have different coordinates.

  The generator supports several "type" parameters to produce a variety of data
  patterns, exposing different corner cases.

  NOTE:
   - Please DO NOT set or parse a random seed parameter.
   - This generator will produce ONE test.
   - We rely on "rnd" from testlib for randomness.
*/

int main(int argc, char* argv[])
{
    registerGen(argc, argv, 1);

    // Parsing parameters with default values where appropriate
    int n = opt<int>("n");
    int m = opt<int>("m");
    int k = opt<int>("k");
    string type = opt<string>("type", "random"); 

    // Basic validation (not strictly required but good for sanity)
    // Ensure 1 <= n*m <= 200
    // Ensure 1 <= k <= min(n*m, 7)
    // We'll assume user input doesn't violate constraints, but you may check if needed.
    if (n <= 0 || m <= 0 || n * m > 200 || k < 1 || k > min(n*m, 7)) {
        cerr << "Invalid parameters: n=" << n << ", m=" << m << ", k=" << k << endl;
        return 1;
    }

    // Create a 2D array to store the number of flowers.
    // We'll fill it depending on the "type".
    vector<vector<int>> garden(n, vector<int>(m, 0));

    // A helper lambda to generate random in [1..1000].
    auto genValue = [&]() {
        return rnd.next(1, 1000);
    };

    // Fill the garden according to "type":
    if (type == "allmin") {
        // All squares have 1 flower
        for (int i = 0; i < n; i++) {
            for (int j = 0; j < m; j++) {
                garden[i][j] = 1;
            }
        }
    } 
    else if (type == "allmax") {
        // All squares have 1000 flowers
        for (int i = 0; i < n; i++) {
            for (int j = 0; j < m; j++) {
                garden[i][j] = 1000;
            }
        }
    }
    else if (type == "random") {
        // Fully random
        for (int i = 0; i < n; i++) {
            for (int j = 0; j < m; j++) {
                garden[i][j] = genValue();
            }
        }
    }
    else {
        // For "corners", "line", "block", or anything else, we'll do a random fill
        // and then handle building squares in a special pattern.
        for (int i = 0; i < n; i++) {
            for (int j = 0; j < m; j++) {
                garden[i][j] = genValue();
            }
        }
    }

    // We'll store the building squares here
    vector<pair<int,int>> buildings;
    buildings.reserve(k);

    // According to "type", choose building squares.
    // Must ensure distinct squares. We'll do it differently for each type.
    // Indices are 1-based for the final output, but we'll pick 0-based internally.
    if (type == "corners") {
        // Up to 4 corners: (0,0), (0,m-1), (n-1,0), (n-1,m-1)
        // If k <= 4, place them in corners first. If k > 4, fill corners, then random for the rest.
        vector<pair<int,int>> corners;
        corners.push_back({0, 0});
        if (m > 1) corners.push_back({0, m-1});
        if (n > 1) corners.push_back({n-1, 0});
        if (n > 1 && m > 1) corners.push_back({n-1, m-1});

        int used = 0;
        for (int c = 0; c < (int)corners.size() && used < k; c++) {
            buildings.push_back(corners[c]);
            used++;
        }
        // If not enough, fill the remainder randomly
        while ((int)buildings.size() < k) {
            int rr = rnd.next(0, n-1);
            int cc = rnd.next(0, m-1);
            if (find(buildings.begin(), buildings.end(), make_pair(rr,cc)) == buildings.end()) {
                buildings.push_back({rr, cc});
            }
        }
    }
    else if (type == "line") {
        // Place building squares in the first row, left to right, then second row, etc.
        // We only do this if it doesn't exceed n*m (which it doesn't: k <= n*m).
        // If k squares don't fit in row 1 alone, continue row by row.
        // Could also do a single row if n=1 or m=2, etc.
        int count = 0;
        for (int i = 0; i < n && count < k; i++) {
            for (int j = 0; j < m && count < k; j++) {
                buildings.push_back({i, j});
                count++;
            }
        }
    }
    else if (type == "block") {
        // Place building squares in top-left block
        // We'll place them in reading order (row by row)
        int count = 0;
        for (int i = 0; i < n && count < k; i++) {
            for (int j = 0; j < m && count < k; j++) {
                buildings.push_back({i, j});
                count++;
            }
        }
    }
    else {
        // "random" or any unknown type => pick k distinct squares at random
        // (If we got here from "random" or leftover type, we handle it similarly.)
        // We'll skip the fill since we already filled the grid above.
        // Just pick k distinct squares randomly.
        set<pair<int,int>> used;
        while ((int)used.size() < k) {
            int rr = rnd.next(0, n-1);
            int cc = rnd.next(0, m-1);
            used.insert({rr, cc});
        }
        for (auto &x: used) {
            buildings.push_back(x);
        }
    }

    // Now we have our matrix of flowers and a set of k building squares.
    // Output format:
    //   line 1: n m k
    //   next n lines: each line has m integers
    //   next k lines: "row col" for building squares (1-based)

    printf("%d %d %d\n", n, m, k);
    for (int i = 0; i < n; i++) {
        for (int j = 0; j < m; j++) {
            printf("%d%c", garden[i][j], (j+1 == m ? '\n' : ' '));
        }
    }
    for (int i = 0; i < k; i++) {
        // Add 1 for 1-based coordinates
        printf("%d %d\n", buildings[i].first + 1, buildings[i].second + 1);
    }

    return 0;
}
\end{lstlisting}

The generator commnads are:

\begin{lstlisting}
# Below are approximately 20 distinct commands to run the generator. 
# Each produces a single test case on stdout.

# 1) Minimum case: 1x1 grid
./gen -n 1 -m 1 -k 1 -type random

# 2) Small case: 1x2 grid, k=1
./gen -n 1 -m 2 -k 1 -type random

# 3) Square grid, small
./gen -n 2 -m 2 -k 2 -type corners

# 4) Another small grid
./gen -n 2 -m 3 -k 2 -type line

# 5) Another small grid with "block"
./gen -n 3 -m 3 -k 3 -type block

# 6) Mixed dimensions, near upper limit in product (10 * 20 = 200)
./gen -n 10 -m 20 -k 3 -type random

# 7) Same n*m=200 but different k
./gen -n 10 -m 20 -k 7 -type corners

# 8) Another large shape, a tall but narrow grid
./gen -n 25 -m 8 -k 5 -type line

# 9) Also near max area
./gen -n 20 -m 10 -k 7 -type block

# 10) All minimal values
./gen -n 5 -m 5 -k 5 -type allmin

# 11) All maximal values
./gen -n 5 -m 5 -k 3 -type allmax

# 12) Very narrow, but tall
./gen -n 100 -m 2 -k 2 -type random

# 13) Another shape with block
./gen -n 8 -m 8 -k 6 -type block

# 14) "line" style with a bit bigger dimension
./gen -n 5 -m 10 -k 4 -type line

# 15) Corners with bigger dimension
./gen -n 4 -m 7 -k 4 -type corners

# 16) Random distribution with k=7
./gen -n 10 -m 10 -k 7 -type random

# 17) Dimensions 4x8 = 32 squares
./gen -n 4 -m 8 -k 3 -type random

# 18) Single row case
./gen -n 1 -m 10 -k 5 -type line

# 19) Single column case
./gen -n 10 -m 1 -k 3 -type block

# 20) Another random large (n*m=200) with max k=7
./gen -n 4 -m 50 -k 7 -type random
\end{lstlisting}

This is also a problem with multiple valid answers. The checker of this problem is:

\begin{lstlisting}[language=c++]
#include "testlib.h"
#include <bits/stdc++.h>
using namespace std;

int n, m, k;
vector<vector<int>> a;
vector<pair<int, int>> important_squares;

int main(int argc, char* argv[]) {
    registerTestlibCmd(argc, argv);

    n = inf.readInt();
    m = inf.readInt();
    k = inf.readInt();

    // Read the garden's flower counts
    a.resize(n, vector<int>(m));
    for (int i = 0; i < n; i++) {
        for (int j = 0; j < m; j++) {
            a[i][j] = inf.readInt();
        }
    }

    // Read the important squares
    set<pair<int, int>> important_set;
    for (int i = 0; i < k; i++) {
        int x = inf.readInt();
        int y = inf.readInt();
        important_squares.emplace_back(x - 1, y - 1);
        important_set.emplace(x - 1, y - 1);
    }

    // Read the jury's answer (minimal total sum)
    int jans = ans.readInt();

    // Read the participant's total sum
    int pans = ouf.readInt();

    // Read the participant's plan
    vector<string> plan(n);
    for (int i = 0; i < n; i++) {
        plan[i] = ouf.readToken();
        if (int(plan[i].length()) != m) {
            quitf(_wa, "Invalid plan: line %d should have length %d, but has length %d", i + 1, m, int(plan[i].length()));
        }
        for (char c : plan[i]) {
            if (c != 'X' && c != '.') {
                quitf(_wa, "Invalid character '%c' in plan at line %d", c, i + 1);
            }
        }
    }

    // Compute the actual total sum over 'X's
    int actual_pans = 0;
    for (int i = 0; i < n; i++) {
        for (int j = 0; j < m; j++) {
            if (plan[i][j] == 'X') {
                actual_pans += a[i][j];
            }
        }
    }

    if (actual_pans != pans) {
        quitf(_wa, "The total sum of dead plants does not match the plan: expected %d, found %d", actual_pans, pans);
    }

    if (pans > jans) {
        quitf(_wa, "Participant's total sum (%d) is greater than minimal total sum (%d)", pans, jans);
    } else if (pans < jans) {
        quitf(_fail, "Participant's total sum (%d) is less than minimal total sum (%d)", pans, jans);
    }

    // Check that all important squares are covered with concrete ('X')
    for (auto [x, y] : important_squares) {
        if (plan[x][y] != 'X') {
            quitf(_wa, "Important square (%d, %d) is not covered with concrete", x + 1, y + 1);
        }
    }

    // Check connectivity between all important squares
    queue<pair<int, int>> q;
    vector<vector<bool>> visited(n, vector<bool>(m, false));
    q.push(important_squares[0]);
    visited[important_squares[0].first][important_squares[0].second] = true;

    // Directions: up, down, left, right
    int dx[] = {-1, 1, 0, 0};
    int dy[] = {0, 0, -1, 1};

    while (!q.empty()) {
        auto [x, y] = q.front();
        q.pop();

        for (int dir = 0; dir < 4; dir++) {
            int nx = x + dx[dir];
            int ny = y + dy[dir];
            if (0 <= nx && nx < n && 0 <= ny && ny < m) {
                if (!visited[nx][ny] && plan[nx][ny] == 'X') {
                    visited[nx][ny] = true;
                    q.emplace(nx, ny);
                }
            }
        }
    }

    // Verify that all important squares are connected
    for (auto [x, y] : important_squares) {
        if (!visited[x][y]) {
            quitf(_wa, "Important square (%d, %d) is not connected to all other important squares", x + 1, y + 1);
        }
    }

    quitf(_ok, "Correct solution with minimal total sum %d", pans);
}
\end{lstlisting}

\subsection{Case Study: Problems with Stronger Test Cases than Official Test Cases}
\label{sec:stronger}

Problem: \href{https://codeforces.com/problemset/problem/392/D}{https://codeforces.com/problemset/problem/392/D}

The following submission passes the official test cases.

\begin{lstlisting}[language=c++]
#include <bits/stdc++.h>
using namespace std;
const int MAXN = 6e5 + 21, inf = 1e7 + 21;
int n, sz;
int A[MAXN], B[MAXN], C[MAXN], a[MAXN], b[MAXN], c[MAXN];
int dp[MAXN], cnt[MAXN];
int bw[MAXN << 2], lazy[MAXN << 2];
struct node {
  int ans, mn, mx;
  node(int a = 0, int b = 0, int c = 0) { mn = a, mx = b, ans = c; }
} seg[MAXN << 2];
inline node MRG(node a, node b) {
  return node(min(a.mn, b.mn), max(a.mx, b.mx), min(a.ans, b.ans));
}
inline void relax(int x, int id, int st) {
  seg[id] = node(lazy[id] = x, x, x <= n ? x + st : 3 * n);
}
inline void shift(int id, int st, int en) {
  if (!(~lazy[id])) return;
  int mid = (st + en) >> 1;
  relax(lazy[id], id << 1, st);
  relax(lazy[id], id << 1 | 1, mid);
  lazy[id] = -1;
}
void build(int id = 1, int st = 0, int en = n + 1) {
  lazy[id] = -1;
  if (en - st == 1) {
    seg[id] = node(dp[st], dp[st], dp[st] + st);
    return;
  }
  int mid = (st + en) >> 1;
  build(id << 1, st, mid);
  build(id << 1 | 1, mid, en);
  seg[id] = MRG(seg[id << 1], seg[id << 1 | 1]);
}
void update(int l, int r, int x, int id = 1, int st = 0, int en = n + 1) {
  if (r <= st || en <= l || seg[id].mn >= x) return;
  if (l <= st && en <= r && seg[id].mx < x) return relax(x, id, st);
  shift(id, st, en);
  int mid = (st + en) >> 1;
  update(l, r, x, id << 1, st, mid);
  update(l, r, x, id << 1 | 1, mid, en);
  seg[id] = MRG(seg[id << 1], seg[id << 1 | 1]);
}
inline void pre() {
  int cur = 0;
  sort(bw, bw + sz);
  sz = unique(bw, bw + sz) - bw;
  fill(A, A + sz, n + 1);
  fill(B, B + sz, n + 1);
  fill(C, C + sz, n + 1);
  for (int i = n; i; i--) {
    a[i] = lower_bound(bw, bw + sz, a[i]) - bw;
    b[i] = lower_bound(bw, bw + sz, b[i]) - bw;
    c[i] = lower_bound(bw, bw + sz, c[i]) - bw;
    A[a[i]] = i;
    B[b[i]] = i;
    C[c[i]] = i;
  }
  for (int i = 1; i <= n; i++) {
    cur += !cnt[a[i]]++;
    cur += !cnt[c[i]]++;
  }
  for (int i = 0, p = n; i <= n; i++) {
    cur += !cnt[b[i]]++;
    while (p && cnt[c[p]] > 1) cnt[c[p--]]--;
    if (cur < sz)
      dp[i] = 3 * n;
    else
      dp[i] = p;
  }
  build();
  return;
}
int main() {
  ios::sync_with_stdio(0), cin.tie(0), cout.tie(0);
  cin >> n;
  for (int i = 1; i <= n; i++) {
    cin >> a[i];
    bw[sz++] = a[i];
  }
  for (int i = 1; i <= n; i++) {
    cin >> b[i];
    bw[sz++] = b[i];
  }
  for (int i = 1; i <= n; i++) {
    cin >> c[i];
    bw[sz++] = c[i];
  }
  pre();
  int ans = n + seg[1].ans;
  for (int i = n; i > 0; i--) {
    if (A[a[i]] == i) {
      update(0, B[a[i]], C[a[i]]);
    }
    ans = min(ans, i + seg[1].ans - 1);
  }
  cout << ans;
  return 0;
}
\end{lstlisting}

The agent system constructed the following input:

\begin{lstlisting}
1
497025789
364691059
954413461
\end{lstlisting}

The correct answer is \texttt{3} but the program outputs \texttt{2}. Therefore, this submission is actually incorrect.
\end{document}